\documentclass[fleqn,usenatbib]{mnras}
\usepackage[T1]{fontenc}

\usepackage{graphicx}	
\usepackage{amsmath}	

\usepackage{xcolor}
\definecolor{violet}{rgb}{0.8,0,1}

\def\revone{}
\usepackage[normalem]{ulem}

\newcommand{\SII}{[S~{\sc ii}]}
\newcommand{\FeII}{Fe~{\sc ii}}
\newcommand{\OI}{O~{\sc i}}
\newcommand{\CaII}{Ca~{\sc ii}}
\newcommand{\SiII}{Si~{\sc ii}}

\newcommand{\OIII}{[O~{\sc iii}]\ }

\newcommand{\NII}{[N~{\sc ii}]\ }
\newcommand{\HII}{H~{\sc ii}}
\newcommand{\HeII}{He~{\sc ii}}
\newcommand{\HeI}{He~{\sc i}\ }
\newcommand{\HI}{H~{\sc i}}

\newcommand{\OIIIHb}{[O~{\sc iii}]/H$\beta$}
\newcommand{\Ha}{H$\alpha$}
\newcommand{\Hb}{H$\beta$}

\title[Diffuse emission in the galaxy  IC~1613]{Nature of the diffuse emission sources in the \HI{} supershell in the galaxy IC~1613}

\author[A. D.  Yarovova  et al.]
{\revone{Anastasiya D. Yarovova,$^{1,2}$}\thanks{E-mail: yaan.ph@gmail.com}
Alexei V. Moiseev,$^{2,3,1}$
Ivan S. Gerasimov,$^{1,2}$
Milica  M.~Vu{\v c}eti{\'c},$^{4}$
\newauthor
Oleg V. Egorov,$^{5,1,2}$ 
Dragana Ili{\'c},$^{4,6}$
Ilya A. Mereminskiy,$^{3}$
Yury V. Pakhomov,$^{7}$
Olga N. Sholukhova$^{2}$
\\
$^{1}$Lomonosov Moscow State University, Sternberg Astronomical Institute, Universitetsky pr. 13, Moscow 119234, Russia\\
$^{2}$Special Astrophysical Observatory, Russian Academy of Sciences, Nizhny Arkhyz 369167, Russia\\
$^{3}$ Space Research Institute of the Russian Academy of Sciences (IKI), Profsoyuznaya 84/32, 117997, Moscow, Russia\\
$^{4}$University of Belgrade, Faculty of Mathematics, Department of Astronomy, Studentski trg 16, 11000 Belgrade, Serbia\\
$^{5}$Astronomisches Rechen-Institut, Zentrum f\"{u}r Astronomie der Universit\"{a}t Heidelberg, M\"{o}nchhofstra\ss e 12-14, D-69120 Heidelberg, Germany\\
$^{6}$Hamburger Sternwarte, Universitat Hamburg, Gojenbergsweg 112, D-21029 Hamburg, Germany\\
$^{7}$Institute of Astronomy of the Russian Academy of Sciences, Pyatnitskaya
str. 48, 119017, Moscow, Russia \\
}

\date{Accepted XXX. Received YYY; in original form ZZZ}

\pubyear{2023}

\begin{document}
\label{firstpage}
\pagerange{\pageref{firstpage}--\pageref{lastpage}}
\maketitle

\begin{abstract}
We present a study of the nearby low-metallicity dwarf galaxy IC 1613, focusing on the search for massive stars and related feedback processes, as well as for faint supernova remnants (SNR) in late stages of evolution. We obtained the deepest images of IC 1613 in the narrow-band \Ha, \HeII{} and \SII{} emission lines and new long-slit spectroscopy observations using several facilities (6-m BTA, 2.5m SAI MSU, and 150RTT telescopes), in combination with the multi-wavelength archival data from MUSE/VLT, VLA, XMM-\textit{Newton}, and {\it Swift}/XRT.
Our deep narrow-band photometry identifies several faint \revone{shells} in the galaxy\revone{, and we} further investigate their physical characteristics with the new long-slit spectroscopy observations and the archival multi-wavelength data. \revone{Based on} energy balance calculations and assumptions about their possible nature, we propose that one of the \revone{shells} is a possible remnant of a supernova explosion. \revone{We study five out of eight Wolf-Rayet (WR) star candidates previously published for this galaxy using the \HeII{} emission line mapping, MUSE/VLT archival spectra, and new long-slit spectra. Our analysis discards the considered WR candidates and finds no new ones.}
\revone{We found P Cyg profiles in \Ha{} line in two stars, } 
which we classify as Luminous Blue Variable (LBV) star candidates. \revone{Overall, the galaxy IC 1613 may have a lower rate of WR star formation than previously suggested}.

\end{abstract}

\maketitle

\section{Introduction}
\label{sec:intro}
Dwarf irregular galaxies provide a unique laboratory for studying the formation of massive stars, their evolution and their interaction with interstellar matter due to the absence of a large-scale spiral structure in combination with a shallow gravitational potential and thick gaseous disc. Despite the small fraction of O-type stars, they emit a huge amount of energy into the surrounding gas through their ionising radiation, powerful stellar winds, and the subsequent supernova explosion. This process creates prominent shell-like structures in the surrounding interstellar medium \citep[ISM, see e.g., ][]{2014prpl.conf..243K, 2019MNRAS.483.2547R, 2021MNRAS.508.2650E,2022MNRAS.517.4968G}. \pagebreak

The joint study of the properties of ISM with stellar population allows for a better understanding of the stellar feedback processes, formation mechanisms, and evolutionary path of massive stars. Of particular interest in the context of these studies are metal-poor galaxies, where both theoretical and observational studies predict changes in how stellar feedback works \citep[see][and references therein]{2001A&A...369..574V, 2023arXiv230710277E}. Star formation and evolution under low metallicity conditions still raise many questions, especially when it comes to the most massive stars, thus more photometric and spectral observations are required \citep[see][and references therein]{2019Galax...7...74N, 2020Galax...8...20W,  2023MNRAS.518.2256Y}. 

\revone{The Local Group irregular galaxy IC 1613 is a perfect testbed for studying the interplay between neutral and ionised hydrogen (\HI{} and \HII) and massive stars in a \revone{low-metallicity} environment}. It is a relatively close \citep[distance 730 kpc according to][]{2001ApJ...550..554D, 2004AJ....127.2031K} and metal-poor \citep[12+log(O/H) = 7.62,][]{2003A&A...401..141L} galaxy. Its stellar content has been first studied in the works of \cite{1971ApJ...166...13S, 1976AJ.....81..743S, 1978ApJS...37..145H}. Further, the structure of the galaxy and the \HII{} regions were described in the works by \cite{1971PUSNO..20d....A} and \cite{1990PASP..102.1245H}. 

\begin{figure*}
\centerline{
\includegraphics[width=0.745\textwidth]{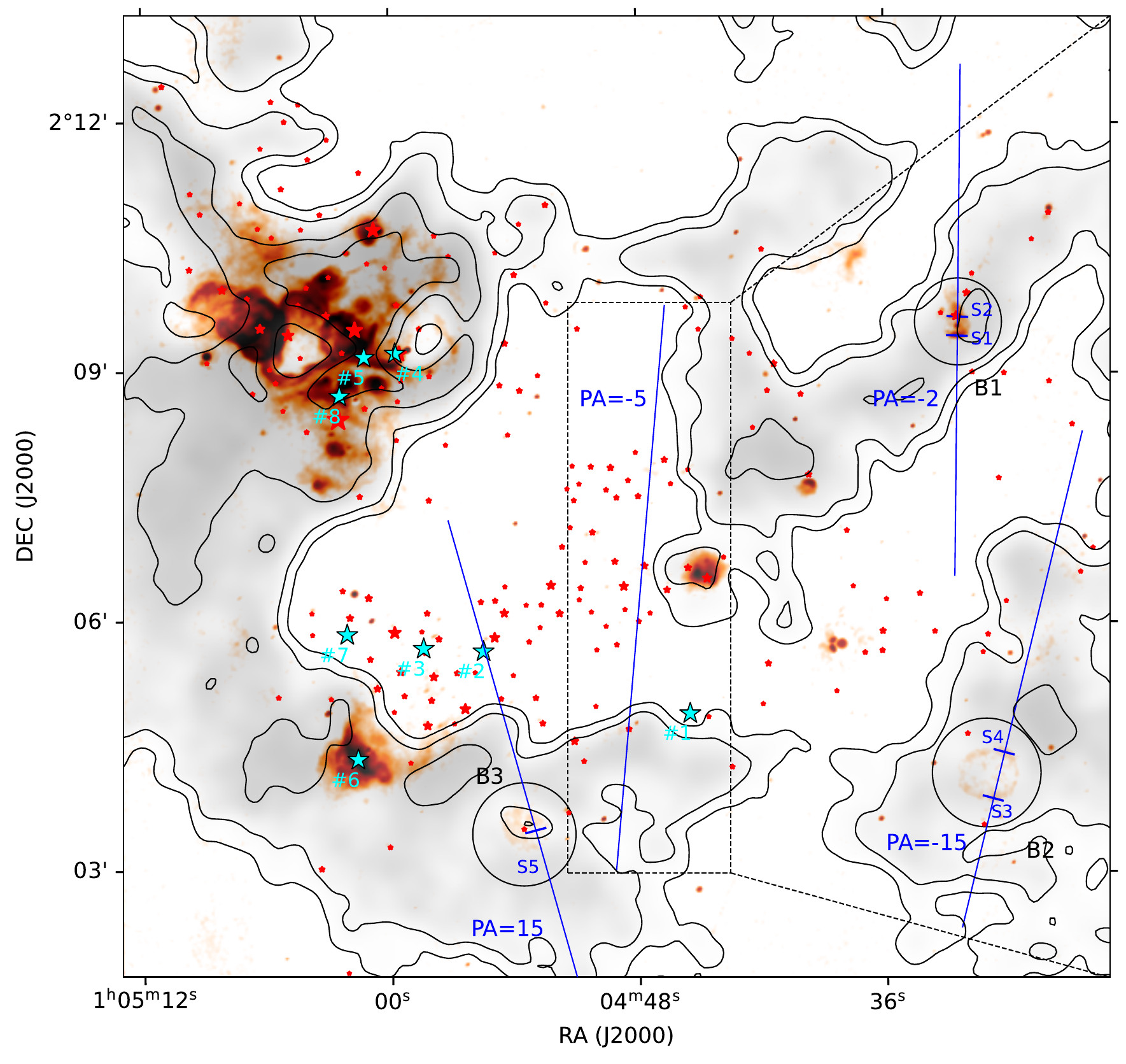}
\includegraphics[width=0.257\textwidth]{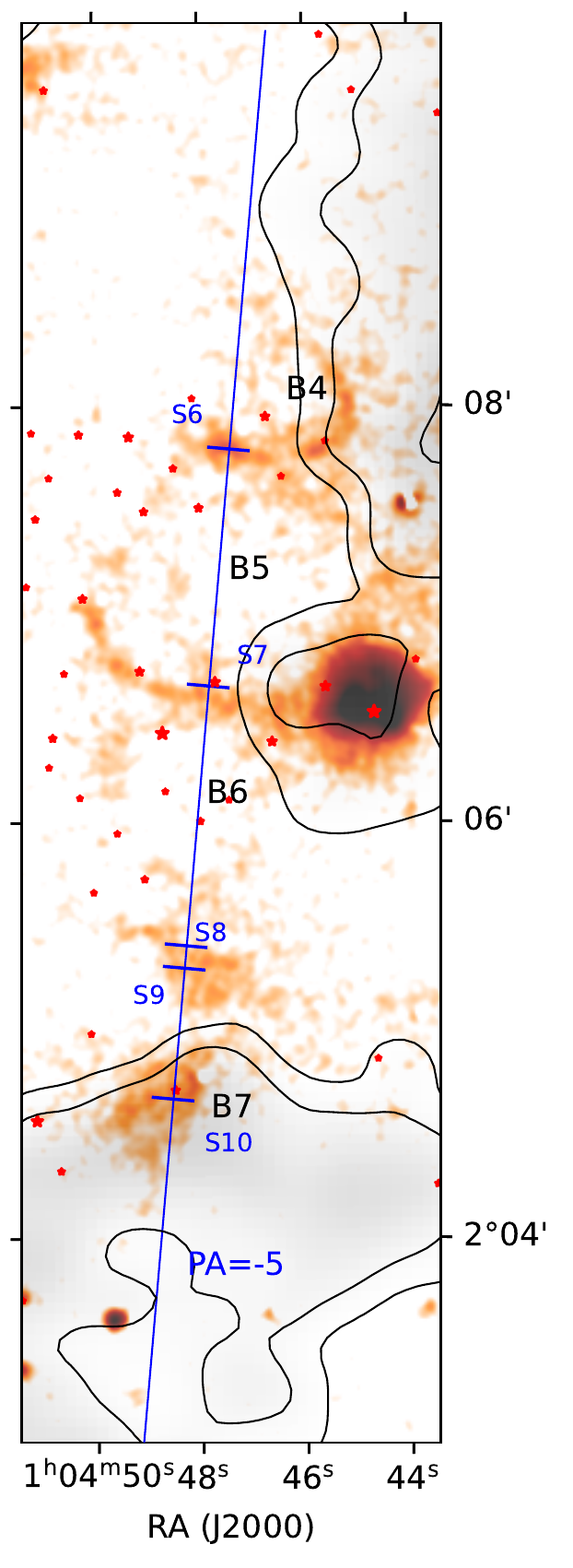}
}
\caption{\revone{\HI{} zero momentum map (grey contours) of IC 1613 from \citet{Lozinskaya2003AstL...29...77L} combined with \Ha{} map (orange colour). The left panel shows a map of the IC 1613 in the \Ha{} line, obtained with the 1.5-m  RTT telescope. The right panel shows a part of the map obtained with the 2.5-m CMO telescope, which reveals especially faint \Ha{} structures. The blue lines on the map indicate the slit positions.  The positions from which the spectra of objects of interest are taken are labelled with S1--S10, while the spectra are shown in Appendix \ref{sec:appendixA} (Fig.~\ref{Spectra_S1-S5} and Fig.~\ref{Spectra_S6-S10}). Black circles mark the faint shells that were analysed in more detail. Red stars mark the positions of OB associations from the \citet{Garcia2009A&A...502.1015G} catalogue. OB associations in shells B1 and B3 are located at their centres, while the B2 shell lacks both OB associations and single O stars. Candidate to Wolf-Rayet stars (cWR) from the work of \citet{ArmandroffMassey1985} are marked by cyan stars with labels. Regions B1-B3 and B7 are located in the dense \HI{} regions, while B4--B6 contact regions of neutral hydrogen only with their edges.}}

\label{fig:Ha_map}
\end{figure*}

Observations of the \revone{brightest star-forming} region in IC 1613 by \citet{Lozinskaya2002ARep...46...16L} revealed the presence of newborn stars at the edges of \HII{} shells \revone{(North-Eastern \Ha\ emitting region on Fig.~\ref{fig:Ha_map})}. Further studies were performed with the \revone{integral field spectrograph} MPFS (Multi Pupil Fibes Spectrograph) to analyse these stars and the surrounding nebulae in detail, leading to the detection of Of star. The findings suggest that shocks have played a significant role in the formation of these shells. 

Using the PUMA scanning Fabry–Perot interferometer, \citet{Valdez-Gutirrez2001A&A...366...35V} performed \Ha{} and \SII{} observations of IC 1613 to analyse the kinematics of ionised gas, showing that the ionised gas is distributed in classical \HII{} regions and in giant shells.
Further Fabry-Perot interferometric studies of the ionised gas kinematics of the central star-forming complex, together with VLA 21-cm radio studies were conducted by \citet{Lozinskaya2003AstL...29...77L}. Three large ($\sim$300 pc) and bright HI supershells were identified in the north-eastern part of the galaxy. They overlap with several smaller \HII{} shells distributed along the walls of the HI supershells and are ionised by young OB associations. The location of the \HII{} shells and the OB associations suggests that sequential star formation was triggered in that region by the expansion of larger neutral hydrogen supershells. 

\citet{Silich2006A&A...448..123S} examined the neutral hydrogen distribution and kinematics in IC 1613, compared them with the ionised gas distribution and stellar content, and also revealed highly inhomogeneous ISM with HI arcs and shells with sizes 200-300 pc in diameter, as well as giant holes and arc-shaped structures. The authors suggest that the observed kpc-scale structure and star formation activity cannot be explained by the multiple supernova hypothesis.

Galaxy IC 1613 hosts one known supernova remnant (SNR) in the central star-forming region in the galaxy. Complete multiwavelength analysis of this SNR, designated as S8, was presented by \citet{1998AJ....116.2328L}. Deep optical images at the KPNO 4-m telescope through narrow-band filters were obtained, as well as VLA and ROSAT  imaging. It was noted that many properties of this SNR are similar to those of the brightest optical SNR in the LMC - N49. More recently, \citet{Schlegel2019AJ....158..137S} described the Chandra observation of S8. According to their observations and previous optical and radio studies of the remnant, it was suggested that the object is a young composite SNR. 
\revone{Subsequent studies of the radio continuum and X-ray emission with VLA and Chandra, respectively, did not uncover any other SNR in IC~1613 \citep{2022AJ....163...66S}.  }

\citet{Garcia2009A&A...502.1015G} \revone{studied the young stellar population of the galaxy and} constructed the most complete catalogue of OB associations, based on the observations with the Wide Field Camera at the Isaac Newton Telescope. They found that the OB associations concentrate in the central region of IC 1613 and confirmed their connection with the \HII{} shells. \revone{The stellar component of IC 1613 has been also recently studied by \citet{2022ApJ...939...28C}, who conducted a spectroscopic study on 14 red supergiant stars in IC 1613.}

\citet{ArmandroffMassey1985} surveyed IC 1613 for \revone{Wolf-Rayet (WR)} stars with narrow-band filters and found 8 WR candidates, 5 of which were labelled as extremely probable. \revone{Later, these stars were spectroscopically studied by \citet{1991AJ....102..927A}. However, for most of the candidates, no certain conclusions were made. Only one WR (WO) star was confirmed, which is still the only known WR star in IC 1613.} The star was discovered by \citet{1982A&A...105..410D}. \revone{Its the most detailed and deepest spectrum was presented by \citet{Tramper2013A&A...559A..72T}, whereas integral-field spectroscopy of the inner part of the surrounding nebula was performed by \citet{Lozinskaya2001ARep...45..417L}}. \revone{Most recently, the stellar component of IC 1613 was studied by \citet{2024arXiv240201631T}, who performed chemo-kinematic analysis for the stars in 3 MUSE/VLT fields for the galaxy. Based on kinematics of individual stars, rotation of the stellar component was detected.}

Our motivation for the present study is to reveal SNRs and massive stars at late evolutionary stages and to characterise the diffuse ionised ISM \revone{beyond the well-studied regions of intense star formation (bright complex of supershells and the nebula around the WO star, located respectively North-East and South-East in Fig.\ref{fig:Ha_map})}. We report our photometric and spectroscopic studies based on the new observations at the 6-m BTA telescope (the Big Telescope Alt-azimuthal, BTA) of the Special Astrophysical Observatory of Russian Academy of Sciences (SAO RAS), 2.5-m telescope of the Caucasus Mountain Observatory (CMO) of Sternberg Astronomical Institute of Moscow State University (SAI MSU), and 1.5-m Russian-Turkish telescope (150RTT) located at TUBITAK National Observatory in Turkey.  These were combined with the archival data from several multi-wavelength facilities: \revone{ Multi-Unit Spectroscopic Explorer (MUSE) at VLT, VLA (\HI{} 21 cm data)}, XMM-{\it Newton}, and {\it Swift}/XRT.

The paper is organised as follows: Section 2 describes the observations and data reduction; Section 3 describes the \revone{ISM morphology and discusses the gas ionisation state and ionisation budget of the studied regions}; in Section 4 we report the results of the search for WR candidates based on \HeII{} narrow-band images; Section 5 discusses stars with H$\alpha$ emission found in MUSE archive spectral data; Section 6 provides the discussion of our finding, and finally Section 7 outlines the main conclusions. 

\section{Observations and data reduction}

\begin{figure}
\includegraphics[scale=0.72]{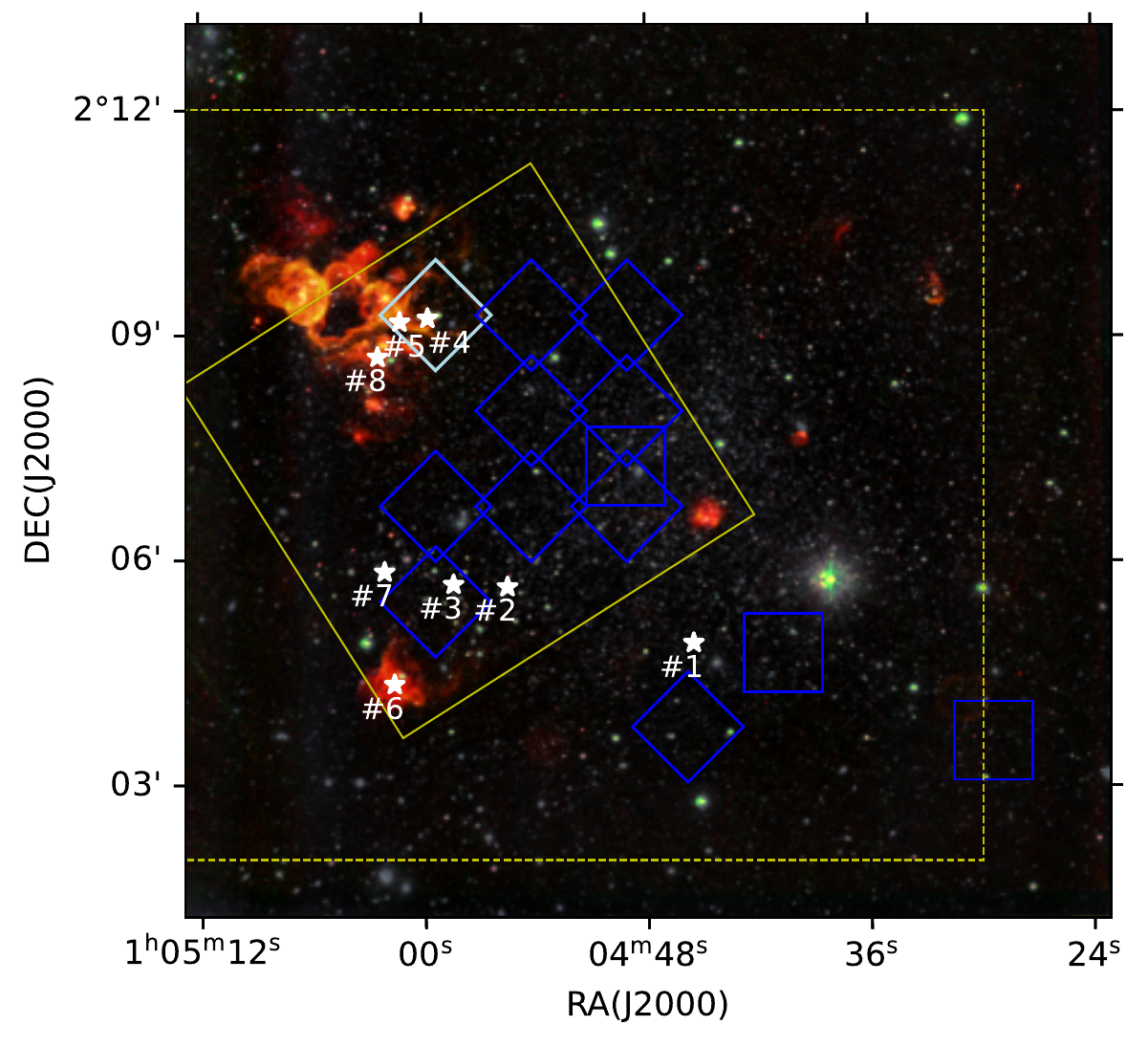}
\caption{Composite \SII{} + \Ha{} + continuum image obtained at the 1.5-m RTT. The yellow dashed square highlights the field of \Ha{} observations from the 2.5-m CMO telescope. The yellow solid line outlines the field mapped with MaNGaL in  \HeII. The positions of the archive MUSE fields are marked with dark and light blue squares.  The latter one is used for BPT diagrams and \Ha{} map calibration. White star symbols indicate the positions of WR star candidates identified in the study of \citet{ArmandroffMassey1985}.}
\label{fig_1}
\end{figure}

\begin{table*}
	\centering
	\caption{ Log of observational data: name of data set, date of observations, total exposure time $T_{exp}$, field of view (FOV), data sampling, angular resolution ($\theta$),   spectral range ($\Delta\lambda$) and  spectral resolution ($FWHM$) }
	\label{t:obs}
	\begin{tabular}{c|c|c|c|c|c|c|c}
	\hline
	Data set              & Date     & $T_{exp}$, s & FOV                             & pixel scale, $''$             & $\theta$, $''$   & $\Delta\lambda$, \AA                 & $\mathrm{FWHM}$, \AA     \\

    \multicolumn{8}{c}{Direct imaging} \\    
        \hline
    TFOSC   & 2015 Sep 12 & 6000 & $13\arcmin\times13\arcmin$ & 0.39 &   1.8       & \Ha{}   & 80 \\
    TFOSC   & 2015 Sep 12 & 6000 & $13\arcmin\times13\arcmin$ & 0.39 &       1.9   & \SII{}   &  54 \\
    TFOSC  &  2015 Sep 12 & 6000 & $13\arcmin\times13\arcmin$ & 0.39 &   1.8       & Continuum (red)   &  350 \\    
    NBI    & 2019 Dec 30 & 3000 & $10\arcmin\times10\arcmin$  & 0.15 &   \revone{2.3}    & \Ha{} & 77 \\   
    NBI    & 2019 Dec 30 & 1800 & $10\arcmin\times10\arcmin$  & 0.15 &    \revone{2.2}   & continuum & 122 \\    
    MaNGaL   & 2020 Nov 12 & 3600 &  $5.6\arcmin\times5.6\arcmin$ & 0.33 &     1.4   &  HeII   & 13 \\
    MaNGaL   &  2020 Nov 12 & 1500 &  $5.6\arcmin\times5.6\arcmin$ & 0.33 &    1.4    &  Continuum (red)   & 13 \\
    MaNGaL   &  2020 Nov 12 & 1500 &  $5.6\arcmin\times5.6\arcmin$ & 0.33 &    1.4    &  Continuum (blue)   & 13 \\
    MaNGaL    &  2020 Nov 14 & 2400 &  $5.6\arcmin\times5.6\arcmin$ & 0.33 &   1.1     &  HeII   & 13 \\
    MaNGaL    &  2020 Nov 14 & 1200 &  $5.6\arcmin\times5.6\arcmin$ & 0.33 &   1.1     &  Continuum (red)   & 13 \\
    MaNGaL    &  2020 Nov 14 & 1200 &  $5.6\arcmin\times5.6\arcmin$ & 0.33 &   1.1     &  Continuum (blue)   & 13 \\
 	\multicolumn{8}{c}{Long-slit spectroscopy}\\
	\hline
	SCORPIO-1  PA=$-2\degr$ & 2020 Sep 24 & 4500 & {$1\arcsec\times6.1\arcmin$}  & 0.36 &  1.7   & 3650-7740  &  12  \\
    SCORPIO-1  PA=$15\degr$ & 2020 Sep 24 & 7200 & {$1\arcsec\times6.1\arcmin$}  & 0.36 &  1.5   & 3650-7740  &  12  \\
    SCORPIO-1  PA=$-15\degr$ & 2020 Sep 24 & 6300 & {$1\arcsec\times6.1\arcmin$}  & 0.36 &  1.1   & 3650-7740  & 12  \\
    SCORPIO-2  PA=$-5\degr$ & 2020 Oct 11 & 5400 & {$1\arcsec\times6.1\arcmin$}   & 0.4 &  1.7   & 3650-7300  & 4.8   \\
    \hline
	\end{tabular}
\end{table*}

\subsection{H${\alpha}$ and SII{} images }
\label{sec:obs_Ha}

Two sets of images were obtained using \Ha{} filters, which differ in the photometric depth, one having moderately larger field of view. Both of the filters also include \NII lines. We estimate contribution of \NII lines to the total flux in the filters as $\leq7\%$, according to the Table \ref{tab:fluxes}. One set \revone{of observations} was obtained in \SII$\lambda6716,6731$ line. 

The deepest image of IC 1613 was obtained with the CMO telescope using the Niels Bohr institute Wide Field CCD Imager  \citep[NBI, main parameters of the camera could be found in][]{NBI2020gbar.conf..127S}. The filter with central wavelength at 6560 \AA\ and bandwidth 77 \AA\ was used to obtain the image.\revone{ FOV of the instrument is $10\arcmin\times10\arcmin$ v.s. $10\arcmin\times8\arcmin$ FOV of the archival \Ha{} image from Little Things Database (\citealt{2012AJ....144..134H})}. 
Data reduction was performed in a standard way including bias subtraction, flat-field correction, correction for non-linearity of the detector, background air-glow emission, and cosmic-ray hits removal. \revone{As the final step, \revone{2$\times$2 binning was applied} to obtain a better S/N value.}

Absolute flux calibration was performed using MUSE data and \Ha{} image before continuum subtraction. For this purpose, we choose a field with a noticeable \Ha{} emission (the field is marked with light blue in Fig.~\ref{fig_1}).  For flux calibration, we firstly convolved MUSE spectra with an \Ha{} filter transmission curve. Then we detected bright stars in the chosen field, both in MUSE data and on \Ha{} image, and performed aperture photometry using DaoStarFinder in Photutils python packages \revone{(see \citealt{1987PASP...99..191S}, \citealt{2023zndo...7804137B})}. The linear regression coefficient between the two data sets was estimated and applied to the \Ha{} image. \revone{The estimated depth (3$\sigma$ above the background level) of this image is $4.5\times10^{-17}\mathrm{erg}\, \mathrm{cm}^{-2}\, \mathrm{s}^{-1}\,  \mathrm{arcsec}^{-2}$, while the archival \Ha{} image from Little Things Database has depth of  $7.0\times10^{-17}\mathrm{erg}\, \mathrm{cm}^{-2}\, \mathrm{s}^{-1}\,  \mathrm{arcsec}^{-2}$}.

Images taken with 150RTT telescope  through \Ha{} and \SII{} narrow-band filters are  lower in depth \revone{than the previously described images}, but have larger field of view ($13\arcmin\times13\arcmin$). The filters width were 80 \AA\ for \Ha{}  and 54 \AA\ for \SII{} line. Data reduction was done using standard procedures in IRAF. Both images were firstly combined using the sigma-clipping method, then the sky was subtracted. We used Astrometry.net\footnote{http://astrometry.net} service \citep{lang2010} to perform astrometric calibration. Background stars were used for removal of continuum radiation.  Images taken through each filter were multiplied with derived coefficients, in order to make stellar fluxes the same  on each filter image. Since standard star images were not taken, flux calibration was performed by using HII regions from \citet{1990PASP..102.1245H}. \revone{The depth of the \SII{} map is $1.5\times10^{-16}\mathrm{erg}\, \mathrm{cm}^{-2}\, \mathrm{s}^{-1}\,  \mathrm{arcsec}^{-2}$.} 
\revone{Log of observations for \Ha{} and \SII{} images is given in Table \ref{t:obs}.}



\subsection{\HeII{} imaging}
\label{sec:obs_phot_HeII}

The image in the \HeII$\lambda4686$ line (which is a characteristic emission in WR stars and other hot sources like SNRe), was  obtained with the 2.5m CMO SAI MSU telescope using the tunable filter photometer  MaNGaL \citep*[Mapper of Narrow Galaxy Lines,][]{Moiseev2020ExA....50..199M}. MaNGaL uses a piezoelectric scanning Fabry-Perot interferometer as a narrow-band filter ($FWHM\approx13$\AA) with a changeable wavelength of a transmission peak.   The filter was subsequently centred on the wavelength that corresponded to the emission line corrected for the systemic velocity of IC1613 and to the continuum shifted in 35\AA\ and 20\AA\   to the red and blue directions correspondingly. The total exposures in each wavelength and the seeing values are listed in Tab.~\ref{t:obs}. The data reduction was done using the standard technique described in detail in previous works \citep*[for example][]{Oparin2020AstBu..75..361O}. The astrometric calibration was done using the  Astrometry.net  service. The mean continuum images were subtracted from the frames in the filter centered to the emission line. The final image shown in Fig.~\ref{fig:HeII}  is a sum of frames obtained during two night in  2017  Nov 12th and 14th. The flux calibration of the  \HeII{} image was based on the standard star observations as described in \citet{Moiseev2020ExA....50..199M}. Because the band of blue continuum also included a possible contribution from C\,{III}$\lambda4650$+N\,{III}$\lambda4634,42$ emission features typical for WC type of WR stars, we checked carefully how the final \HeII{}  map depends on which type of continuum image is being subtracted: blue, red or their arithmetic average. The results were  similar  in all three versions, therefore below we used only   mean continuum subtraction, because it produces better S/N ratio.  \revone{The depth of the \HeII{}  map is $6\times10^{-17}$~erg~s$^{-1}$~cm$^{-2}$~arcsec$^{-2}$.}

Fig.~\ref{fig_1} shows composite \Ha, \SII{} and continuum image taken with 150RTT telescope as well as positions of \Ha{} fields from 2.5-m CMO telescope, \HeII{} image and MUSE fields. 

\subsection{Long-slit spectroscopy}
\label{sec:obs_longslit}

Long-slit spectra were obtained with the 6-m BTA telescope using the SCORPIO-1 \citep{2011BaltA..20..363A} and SCORPIO-2 \citep{2017AstBu..72..458A} multimode focal reducers. The description of the observation parameters is presented in the Table \ref{t:obs}. The first three spectra were obtained in September 2020 with   SCORPIO-1 and the VPHG550G  grating  in the spectral range 3650-7740 \AA. The fourth spectrum was obtained in December 2020 with   SCORPIO-2   and the VPHG1200@540 grating  in the spectral range 3650-7300 \AA. In all cases, the slit width was 1~\arcsec. 
Position of the spectrograph's slit on the \Ha{} image is shown in Fig.~\ref{fig:Ha_map}.

Data processing was performed in a standard way using an IDL-based program, as described in our previous papers \citep[for example][]{Egorov2018}.

To calibrate the long-slit spectra to the absolute intensity, we used the spectra of spectrophotometric standard stars obtained on the same nights, located at a similar zenith distance immediately after or before the object observations.

For each of the studied regions we obtained one-dimensional spectra by summing the signal in apertures corresponding to 2--3 seeing values, visually making sure that the signal from the objects completely falls into the aperture. To measure the emission line fluxes, we used our own code running on {\sc python}. We applied the Gaussian approximation to measure the integral fluxes in the lines of the studied areas.

Since the studied \Ha{} regions are faint, \Ha/\Hb{} ratio shows variation for different summation limits, which means that we cannot accurately perform the standard reddening correction method based on the Balmer decrement. We made a correction for reddening in Milky Way, assuming that IC 1613 is a metal-poor galaxy, so noticeable absorption in it is not expected. We used value $\mathrm{E}_{\mathrm{B-V}} = 0.036$ from \citet{Planck2013} obtained via the `dustmaps' python software \citep{2018JOSS....3..695M}. Then we utilized the \cite{Cardelli1989} curve parametrized by \cite{Fitzpatrick1999}. The reddening-corrected spectra are plotted on Fig.~\ref{Spectra_S1-S5} and Fig.~\ref{Spectra_S6-S10}. All the measured fluxes are given in Table \ref{tab:fluxes}. 
 
The uncertainties associated with the measured flux values were determined using the Monte Carlo method. For certain emission lines that were required for the construction of diagnostic diagrams (Sec.~\ref{sec:3_ion_bud}), but not discernible in the spectra, an upper limit was estimated from the noise level: \revone{$$\mathrm{\sigma_l = 3\sigma_cFWHM\sqrt{\pi/(4\ln2)}.}$$ Here $\sigma_c$ represents the standard deviation in the continuum near the studied line and $\mathrm{FWHM}$ is the full width at half maximum measured for the visible spectral lines (\AA).}

\subsection{Archival data}

We used archival data from the Multi-Unit Spectroscopic Explorer (MUSE; \citeauthor{Bacon2010}~\citeyear{Bacon2010}) for the study of WR candidates and \Ha-emission stars in IC 1613. MUSE data were also used for calibration of \Ha{} field from 2.5-m telescope. The data are available for 13 fields covering IC 1613 as a part of the programs PI: Bian; 105.20GY.001 and PI: Battaglia; 097.B-0373(A). 
\revone{The observations were performed without adaptive optics.} Spectra cover wavelengths from 4750 to 9350\AA\  with spectral resolving power at central wavelength $R=3027$. \revone{Seeing varied from} 0.7 to 1.1$\arcsec$ for different cubes.  
For our analysis, we used \revone{phase 3 data cubes (downloaded from ESO science archive)}, which have undergone a standard reduction using the pipeline described by \cite{Weilbacher2020}. 

After checking the coverage of our objects of interest (see Fig.~\ref{fig_1} for WR star candidates from \citet{ArmandroffMassey1985}), we found out that three of eight WR candidates have been covered by these fields, and extracted the MUSE spectra of these stars. Our analysis of stars with \Ha{}  emission in the MUSE fields is given in Section \ref{emission_stars_fromMUSE}. 

For the analysis of the nature of \revone{faint} \revone{shells} we also used 21cm radio observations from VLA. For a detailed description of the radio data see the work of \citet{2003AstL...29...77L}.  We investigate here the 21cm zero-momentum map translated to the neutral hydrogen surface density. The beam size of $7.4\times7.0$ arcsec corresponds to a linear resolution of about 26 pc.

Finally, we explored the XMM-{\it Newton} \revone{\citep{2001A&A...365L...1J} and {\it Swift}/XRT \citep{2004ApJ...611.1005G,2005SSRv..120..165B}} archival data to support the analysis of the \revone{shells} origin.

\section{The diffuse ionised  shells}
\label{sec:diff_ion_shells}

\revone{Our narrow-band imaging reveals} faint ionised nebulae and chains of filaments on the opposite side from the main star-forming region of IC 1613.  \revone{Faint shells studied in the present work are denoted as B1-B7 on Fig.\ref{fig:Ha_map}. Several shells are visible in the archival \Ha{} images (B3, B5 and B7). However, thanks to better sensitivity and larger FOV of our data, we were able to detect other shells}. These structures have not been previously \revone{spectroscopically} studied, therefore we performed slit spectroscopy of several faint \Ha{} regions. Four slit positions (with PA = 15, PA = -5, PA = -2, PA = -15) crossed three \Ha{} \revone{shells} (B1, B2, B3, marked in Fig.~\ref{fig:Ha_map}) and a chain of faint filaments and \revone{shells} B4--B7 (shown in the right panel of Fig.~\ref{fig:Ha_map}). \revone{According to Fig.~\ref{fig_1}, four of the studied \revone{shells} (B1, B2, B3, B7) are located in the areas of dense \HI{} regions, while B4--B6 \revone{shells} lie near the edges of the regions of neutral hydrogen.}

For further analysis, we extracted the spectra of  10 regions marked \revone{with $S_1$ to $S_{10}$} on Fig.~\ref{fig:Ha_map}.
The first two spectra were extracted from the regions $S_1$ and $S_2$, which correspond to the edge of the \revone{shell} B1 in Fig.~\ref{fig:Ha_map}. This \revone{shell}  has a size of 71 pc. According to \cite{Garcia2009A&A...502.1015G}, this region hosts an OB association that contains 18 stars, 17 of which are of O and B types.

The regions $S_3$ and $S_4$ are located at the edges of \revone{shell} B2. In Sec. \ref{sec:3_ion_bud} we discuss the possible source of ionisation of this \Ha{} region, as we do not see  OB association inside this \revone{shell}. The size of the \revone{shell} is \revone{125 pc}.

Spectrum $S_5$ was extracted from the \revone{106} pc-size nebula (B3 on Fig.~\ref{fig:Ha_map}), with homogeneous structure containing small OB association (4 OB stars according to \citealt{Garcia2009A&A...502.1015G}).

Slit with PA=-5 crosses a chain of \revone{shells} (B4, B5, B6), a single \revone{shell} B7 and a gaseous structure between them. \revone{B4-B7 shells have sizes 113, 247, 158 and 66 pc, respectively.} \revone{Shells} B4, B6, and B7 are possibly connected with OB associations inside. \revone{Central part of shell B5} does not overlap with any OB association, but it can be noticed that young stellar population follows the ring of ionised gas. However, despite the fact that OB associations follow the edges of the \revone{shell} B5, if we consider individual O stars from the complete \citet{Garcia2009A&A...502.1015G} catalog\footnote{from private communication}, such clear spatial matching is not present. The observed distribution of OB associations may indicate a wave of secondary star formation on the \revone{shell} walls, where the density of ISM is higher.

\subsection{Gas ionisation state}
\label{sub:gas_ion_st}
To study gas ionisation state in the faint \revone{shells} we performed BPT-diagram diagnostics \citep*{BPT}  based on \OIII/\Hb, versus \NII/\Ha{} and \SII/\Ha. We take lines {separating photo- and shock ionisation regions from the \revone{classical} works of \citet{Kewley2001} and \citet{Kauffmann2003}, see Fig.~\ref{BPT}. In particular, regions of photoionisation typically lie below the both separating curves, while SNRs lie above (nebula between the two dividing lines in the diagram can be ionised by combination of radiation and shocks). The above division was obtained for the ISM metallicity close to the Solar one. Depending on the metallicity of the analyzed gas, it is expected that the dividing lines should shift (move to the left for lower metallicity). According to  \citet{2003A&A...401..141L}, metallicity for IC 1613 is $\mathrm{12+\log(O/H) = 7.62}$. To estimate the effect of low metallicity on separating lines on BPT diagrams shown on Fig.~\ref{BPT}  we additionally plot the BOND models of photoionisation from \cite{Vale2016} with fixed metallicity $\rm 12 + \log(O/H) = 8.0$ and $\mathrm{N/O} = -1.5$, which is typical for galaxies with such metallicity \cite[see Fig. 6 in][]{Berg2012}. We also show line ratios for bright \Ha{} regions from MUSE data on the BPT diagrams. Assuming that they are normal \HII{} regions, they provide a reference sample for other nebulae on BPT diagrams for IC1613 metallicity. \revone{We also compare the position of the faint \revone{shells} on BPT diagram with the model grid from \cite{Vale2016} for low gas-phase metallicity. Note that line ratios for bright \Ha{} regions from MUSE data are in good agreement with the models from \cite{Vale2016}, which can be seen in Fig. \ref{BPT} }

Most of the obtained long-slit spectra (see Appendix.~\ref{sec:appendixA}, Fig.~\ref{Spectra_S1-S5} and Fig.~\ref{Spectra_S6-S10}) show bright \SII{}\ lines, while the \NII{} lines are often not visible. Also, not all of the spectra show sufficient flux in \OIII$\lambda5007$ line. For those spectra where one of the fluxes in the diagnostic lines is not detected, we estimated its upper limit (indicated by an arrow on the graph).

None of the points on the BPT diagrams (Fig.~\ref{BPT}) locate far outside  the separating lines, \revone{although point S2 is the farthest from the models of \citet{Vale2016}. From the diagram, we can conclude that shock waves could contribute to the ionisation of one of the \revone{shells} (B1 with corresponding spectra S2).} The rest of the regions do not show evidence of shock ionisation. 

\Hb{} line in the $S_8$ spectrum cannot be measured, so we estimated it on the assumption that \Ha/\Hb= 2.86, which is a typical value for a gas at a temperature of 10,000 K \citep{Osterbrock2006}. Under this assumption, the \mbox{[O~\textsc{iii}]/\Hb{}} flux ratio value is not less than 8.8. Such a high ratio of the fluxes can be explained by a planetary nebula that crossed the field of the slit. \revone{For these objects, it is typical to have significant \OIIIHb\ ratio (see \citealt{Osterbrock2006}).}

\begin{table*}
\centering
\caption{Emission line fluxes relative to \Hb{} line for the studied regions in IC 1613. For lines not exceeding the noise limit, we have estimated the upper flux limit (see detailed description in \ref{sub:gas_ion_st}).}
\hspace{1.0cm}
\label{tab:fluxes}
\begin{tabular}
{|l|l|l|l|l|l|l|}
\hline
\# & \Ha/\Hb{} &{[}OIII{]} $\lambda$5007 & {[}OI{]} $\lambda$6300 & {[}NII{]} $\lambda$6583 & {[}SII{]} $\lambda$6716 & {[}SII{]} $\lambda$6731\\ 
\hline
  $S_1$& 2.8 $\pm$ 0.1 &   0.82 $\pm$ 0.05               &  0.30 $\pm$ 0.03 &  0.130 $\pm$ 0.018 &  0.66 $\pm$ 0.04  & 0.45 $\pm$ 0.04  \\ 
  $S_2$ & 2.91 $\pm$ 0.17  & 2.64 $\pm$ 0.15   & &   & 0.40 $\pm$ 0.06 & 0.37 $\pm$ 0.05 \\ 
  $S_3$ & 3.1 $\pm$ 0.4 &                  \revone{<0.6}                        &               &   & 0.85 $\pm$ 0.17  &  0.57 $\pm$ 0.13 \\ 
  $S_4$ & 3.0 $\pm$ 0.3 &                    \revone{<0.4}                     &               &    &  0.73 $\pm$ 0.12  &  0.49 $\pm$ 0.09 \\ 
  $S_5$ & 2.9 $\pm$ 0.4 &  0.95 $\pm$ 0.19                   &   &  & 0.30 $\pm$ 0.08 & 0.20 $\pm$ 0.07                \\
  $S_6$& 2.9 $\pm$ 0.6 &   0.60 $\pm$ 0.26                  &   & 0.18 $\pm$  0.08 & 0.73 $\pm$ 0.18 &  0.39 $\pm$ 0.10         \\ 
  $S_7$& 2.6 $\pm$ 0.7 &  \revone{<0.4}   &                    &  & 0.53 $\pm$ 0.18 &    0.41 $\pm$ 0.14             \\
  $S_8$ &  \multicolumn{1}{c|}{-}       &  >8.8 &              &   & &                 \\
  
  $S_9$ & 3.7 $\pm$ 2.6 &  \revone{<0.5}   &                   & & 0.7 $\pm$ 0.6  &      0.4 $\pm$ 0.3           \\
  
  $S_{10}$ & 3.5 $\pm$ 1.5 &    \revone{<0.4}   &                  & & 0.7 $\pm$ 0.4  &    0.42 $\pm$ 0.22             \\
\hline  
\end{tabular}
\end{table*}

\begin{figure*}
\includegraphics[scale=0.7]{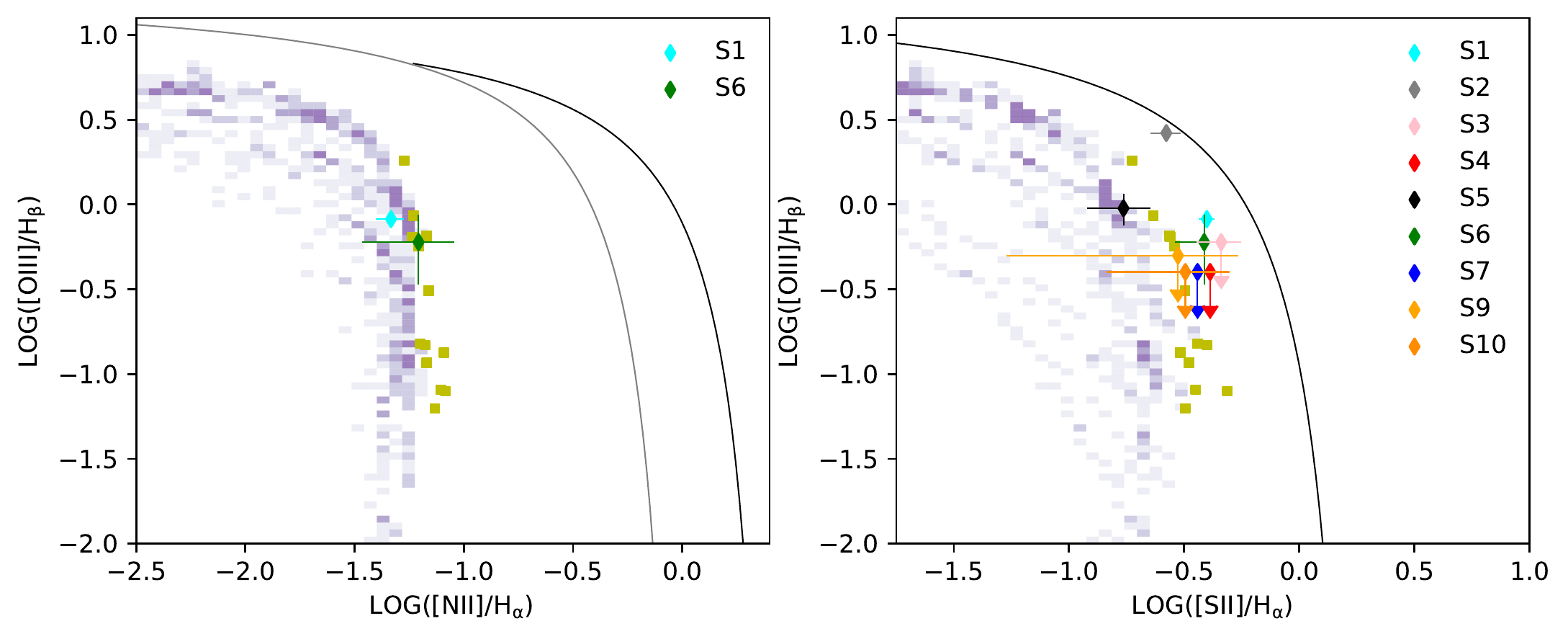}
\caption{The diagnostic BPT diagrams showing the line ratios of \OIII{}/\Hb, versus \NII/\Ha{} (left-hand panel) and \SII/\Ha{} (right-hand panel) for the studied faint ionised  regions listed in the Table \ref{tab:fluxes}. Corresponding spectra are plotted on Figs.~\ref{Spectra_S1-S5} and \ref{Spectra_S6-S10}. Diagrams include black and grey curves from \citet{Kewley2001} and \citet{Kauffmann2003}, that serve as boundary between photoionised regions (located to the left of the lines) and regions with high contributions of other excitation mechanisms. The models with fixed $12 + \log(\mathrm{O/H}) = 8.0$  and $\mathrm{N/O} = -1.5$ from \citet{Vale2016} are overlaid as histogram.  Yellow points correspond to MUSE data for the bright \Ha{} region, presumably dominated by photoionisation (the corresponding region is marked with light blue square on Fig.~\ref{fig_1}).
}
\label{BPT}
\end{figure*}

\subsection{The ionisation budget}
\label{sec:3_ion_bud}
The flux in the \Ha{} line from faint \revone{shells} was estimated using the 2.5-m telescope \Ha{} map, \revone{as it has better sensitivity and more reliable flux calibration than the map from 1.5-m telescope}. As the \revone{shells} B2 and B3 are partially outside the image field, we estimated their fluxes extrapolating the flux distribution from the visible parts. We assumed circular geometric shapes and homogeneous brightness of \revone{shells}. The assumption is based on the image from the 1.5-m telescope, \revone{which has larger FOV, making \revone{shells} B2 and B3 completely visible)}. Then we multiplied the measured partial flux from the map from the 2.5-m telescope by the corresponding coefficient. Flux errors were estimated using the standard deviation of the background values and do not include systematic errors. The background was subtracted for each region individually using patches from neighbouring regions.

To compare the \Ha{} fluxes of the \revone{shells} \revone{with the hydrogen-ionizing Lyman continuum ($L_{y}C$) photons (with $\lambda < 912$~\AA)} produced by young clusters at their centres, we used the relation between the number of $L_{y}C$ photons ($\mathrm{Q_{H\alpha}}$) and $\mathrm{M_V}$ fluxes of O stars, as established by \citet{2005A&A...436.1049M}. To obtain the $\mathrm{M_V}$ fluxes, we referred to the catalogue of stellar photometry from \citet{Garcia2009A&A...502.1015G}. We estimated amount of $L_{y}C$ photons required for ionisation of studied \revone{shell} ($\mathrm{Q_{H\alpha}}$) using the relation from \cite{Osterbrock2006}: $Q^0_\mathrm{H\alpha}\simeq \frac{L\mathrm(H\alpha)}{0.45 h\nu}$.

The B1 region containing the OB association of 17 OB stars inside demonstrates an excess of ionising photons produced by massive stars in comparison with photons, required for ionisation of the \revone{shell} ($\mathrm{Q_{H\alpha}}$=(2.91 $\pm$ 0.01)$\times10^{-48}s^{-1}$ vs. $\mathrm{Q_{M_V}=54\times10^{-48}s^{-1}}$), indicating a potential leakage of ionising photons into the ISM. The B3 \revone{shell} also contains an OB association in the centre, but it is much smaller (4 OB stars). This association contains only one star with $M_\mathrm{V}<-3.8$, therefore we estimated the number of ionizing photons from it. According to our calculations, the number of ionising photons from this star is $\mathrm{Q_{M_v}=0.9\times10^{-48}s^{-1}}$ vs. $\mathrm{Q_{H\alpha}}$=(2.27 $\pm$ 0.01)$\times10^{-48}s^{-1}$, which means that radiation from the star is not sufficient to ionise the nebula. 
The area B2 does not contain OB-associations nor single O stars. The brightest star in this \revone{shell} has an absolute magnitude of $M_\mathrm{V}$ = -3.2, whereas other stars are significantly less luminous. Given that the ionisation of this \revone{shell} requires a comparable number of photons as for the other \revone{shells} under consideration ($\mathrm{Q_{H\alpha}}$=(3.06 $\pm$ 0.01)$\times10^{-48}s^{-1}$), we cannot explain its ionisation by radiation from massive stars.

We did not evaluate the energy balance for \revone{shells} B4--B6, since there is no clear connection between these \HII{} \revone{shells} and OB associations capable of ionizing them. Inside the B7 \revone{shell} one OB association is located, containing 6 OB stars. For this region, $\mathrm{Q_{H\alpha}}$=(1.37 $\pm$ 0.01)$\times10^{-48}s^{-1}$. The region contains one very bright star with $M_\mathrm{V}$=-6.4, which can ionise the whole \revone{shell}, while other stars are less luminous having $M_\mathrm{V}$>-3.

\subsection{SNR candidates}
Among the faint structures found, only the B2 \revone{shell} can be considered as a possible SNR. Based on the analysis of the \citet{Garcia2009A&A...502.1015G} catalogue, no O stars are observed within this region. Nevertheless, the cumulative \Ha{} flux from  B2 is found to be comparable to that of nearby B1 and B3 regions. Moreover, its spherically symmetrical shape possibly suggests the presence of the central source of ionisation and/or gas outflow.
The absence of discernible objects within the \revone{shell} raises the possibility that it may be an SNR. However, we do not see signatures of shocks on BPT diagrams.\\

We also investigated the region based on radio observations from the work of \citet{2022AJ....163...66S}. However, these observations, which encompassed the entire galaxy with a resolution of $5''$, did not reveal any significant radio sources within the B2 region with the limiting sensitivity in flux $4.3\times10^{-15}$ erg cm$^{-2}$ s$^{-1}$ corresponded to the unabsorbed luminosity $7.8\times10^{34}$ erg cm$^{-2}$. Additionally, Chandra X-ray observations from the same study did not cover the B2 region, limiting our ability to assess any potential X-ray emissions. Although not covered by archival {\it Chandra} imaging, the B2 \revone{shell} was covered by both XMM-{\it Newton} and {\it Swift}/XRT observations. Both observatories detected a single point source inside a \revone{shell} - designated as 4XMM J010431.4+020409 \citep{webb20_xmmdr4} / 2SXPS J010431.3+020410 \citep{evans20}, located at $\approx 10\arcsec$ offset from the centre of the \revone{shell}. 
We extracted source spectrum from XMM-{\it Newton} observation 0781200401 with XMM SASv20, using 15$\arcsec$ and 40$\arcsec$ circular regions for source and background, correspondingly. Background region was located on the same EPIC-pn chip, as close to the source position in order to account for possible contribution of diffuse emission from galaxy. Given the source weakness ($F_{\mathrm{X}} = 3.3\times10^{-14}$ erg cm$^{-2}$ s$^{-1}$ in 0.5-10 keV band) we chose to use a simple absorbed powerlaw for spectral analysis ({\sc tbabs*powerlaw}). Measured absorption column density $N_{H} = 3^{+4}_{-2} \times 10^{22}$ cm$^{-2}$ exceeds the Galactic contribution along the line of sight $N_{H, Gal} = 5\times10^{20}$ cm$^{-2}$ \citep{HI4PI}, while the powerlaw index is poorly constrained $\Gamma = 2.3^{+1.8}_{-1.2}$. Assuming that the source is located in IC 1613  we can estimate the intrinsic luminosity to be $L_{X}\approx 6\times10^{36}$ erg s$^{-1}$, which is typical for X-ray binaries or bright pulsars/pulsar wind nebulae. 

There are until now several known X-ray binaries located in their parental SNRs \citep{maitra21}, all of them are hosting hot and massive stars. Yet, as it was already mentioned, there are no bright optical stars inside the \revone{shell}, so the X-ray binary scenario is unlikely. We have also checked the $H\alpha$ image at the position of the source but have not found any emission knots. Lack of the detection in radio bands also makes it improbable that the source is a young pulsar. We tentatively propose that 4XMM J010431.4+020409 may be a background AGN, although additional radio/NIR/X-ray observations are needed to prove this hypothesis. 

In light of these findings, the origin and ionisation mechanism of the \revone{shell} structure in Region B2 remain controversial.

\section{The WR candidates}

\label{sec:WR_candidates}
\subsection{WR candidates from \HeII{} images}
\citet{ArmandroffMassey1985}  performed a search of WR stars in three areas in IC1613 using narrow-band (FWHM=55\AA) filters centred on the spectral emission features specific to WC type (C\,{III}$\lambda4650$+N\,{III}$\lambda4634,42$)  and WN type (\HeII$\lambda4686$)  stars and on the red continuum ($\approx\lambda4750$). They found 8 candidates emit in these lines: six WN stars (the star \#8 was further classified as SNR, see Sec.~\ref{sec:intro}), one WC star (\#4) and already known star \#6 that was classified as a very rare WO-type star \citep{Davidson1982PASP...94..634D,Garnett1991ApJ...373..458G}. \revone{In the subsequent work by \citet{1991AJ....102..927A}, cWR \#6 was confirmed and cWR \#4 rejected based on spectral data. It was also shown, that cWR \#8 is probably a young supernova remnant. For other cWR candidates, their data were not deep enough to draw reliable conclusions about the nature of the objects.} These finding  motivated as to use IC1613 as a test target for the  first MaNGaL observations  in \HeII{} line. We have obtained a  deep image collecting in total 6000/5000 s in the emission  line/continuum at the 2.5m telescope  with a good seeing ($<1.5''$). Moreover, a tunable filter photometry  provides very accurate subtraction of underlining continuum comparing with  `a classic' imaging in narrow-band filters \citep{Moiseev2020ExA....50..199M}. 
It allowed us to get a typical value of RMS error in the continuum-free image (Fig.~\ref{fig:HeII}, left) $6.3\times10^{-18}$~erg~s$^{-1}$~cm$^{-2}$ per pixel that corresponds to  detection with $S/N>3$ of   star-like sources with an integrated \HeII{} flux brighter than $5\times10^{-17}$~erg~s$^{-1}$~cm$^{-2}$ and  extended structures with a  surface brightness larger than  $6\times10^{-17}$~erg~s$^{-1}$~cm$^{-2}$~arcsec$^{-2}$ \revone{(see Sec.~\ref{sec:obs_phot_HeII})}. 

Fig.~\ref{fig:HeII} clearly demonstrates that only two \HeII{} emitters were found out of 7 candidates from the list of \citet{ArmandroffMassey1985} in the MaNGaL field of view. The zoomed fields around WR candidates are shown in Fig.~\ref{fig:HeII_WR} together with DESI Legacy survey images. All candidates  spanning  magnitudes in blue continuum from 19.8 to 20.8 are detected with a high $S/N$ in MaNGaL images, however   \HeII{} emission  appeared above the detection level only in the star \#6 (WO) and star \#8 (SNR).  To verify this conclusion we evaluated  and indicated in Fig.~\ref{fig:HeII_WR}  $\Delta m$ --- the difference between aperture  magnitudes in the filter centered on the emission line and continuum   \citep[$WN-CT$ in terms from][]{ArmandroffMassey1985}. This value is  negative for a star with \HeII{} emission (WO and SNR), whereas $\Delta m\ge 0$ in all other WR candidates. In addition to the already known  \HeII{} emitters (\#6 and \#8) we found only one additional emission source marked by us according it's DESI Legacy survey coordinates as  J010457.6+020953. It  seems like a background distant galaxy (Fig.~\ref{fig:HeII_WR}).

\begin{figure*}
\centerline{
\includegraphics[height=0.6\textwidth]{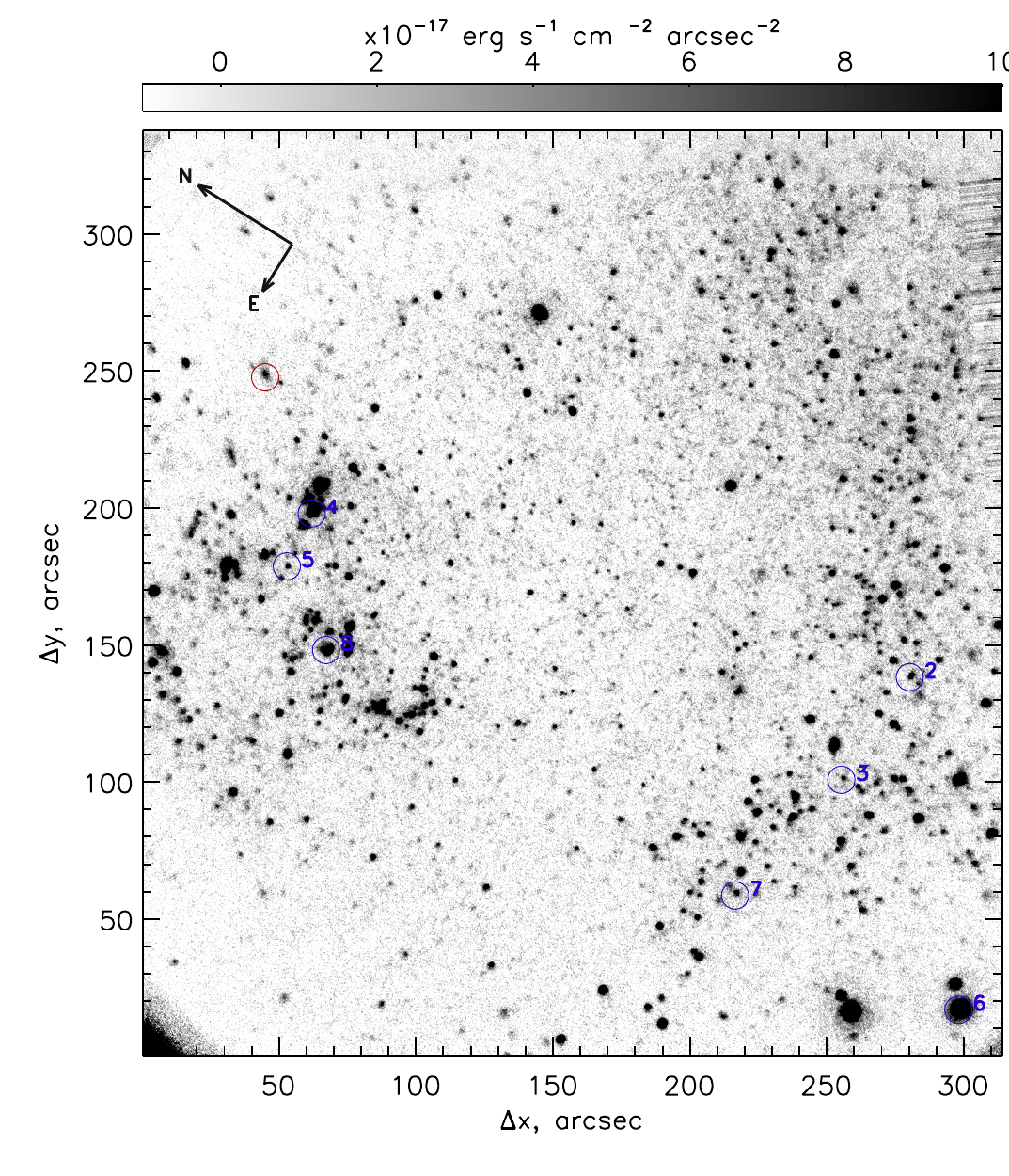}
\includegraphics[height=0.6\textwidth]{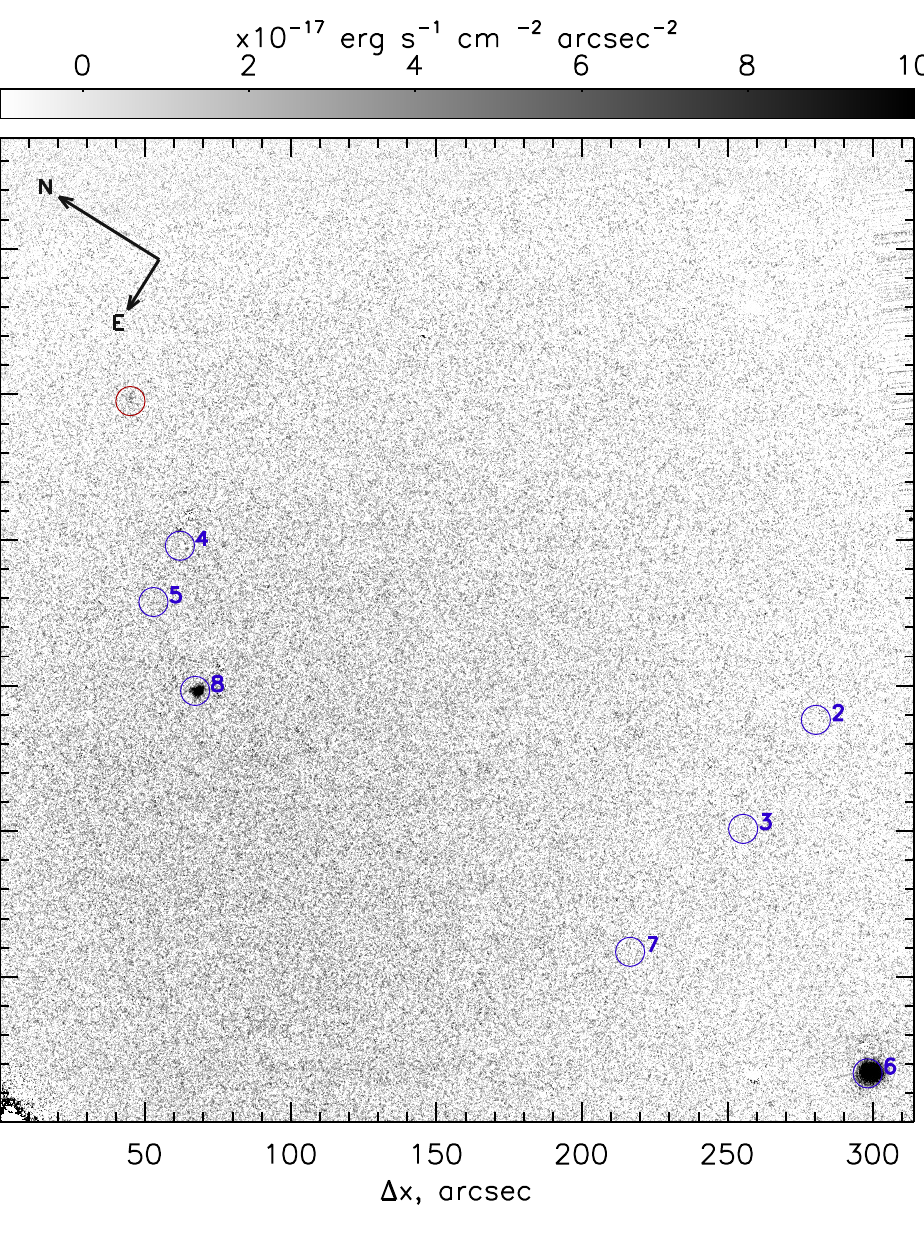}
}
\caption{MaNGaL images in the \HeII{} emission line before (left) and after (right) continuum subtraction. The blue circles (10 arcsec in diameter) centered on the WR-candidate according   \citet{ArmandroffMassey1985}, the red circle marks a background galaxy J010457.6+020953.}
\label{fig:HeII}
\end{figure*}

\begin{figure*}
\centerline{
\includegraphics[width=0.5\textwidth]{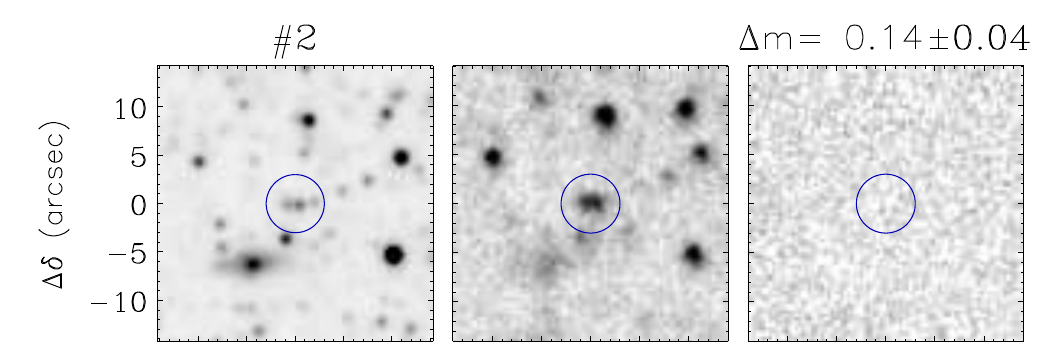}
\includegraphics[width=0.5\textwidth]{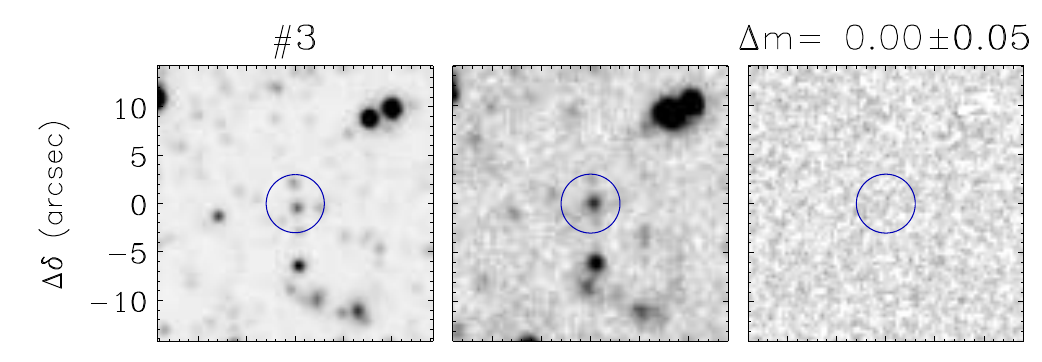}
}
\centerline{
\includegraphics[width=0.5\textwidth]{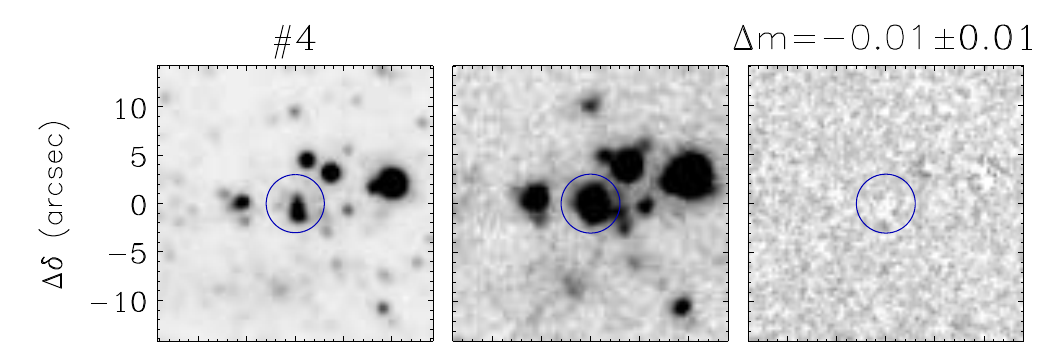}
\includegraphics[width=0.5\textwidth]{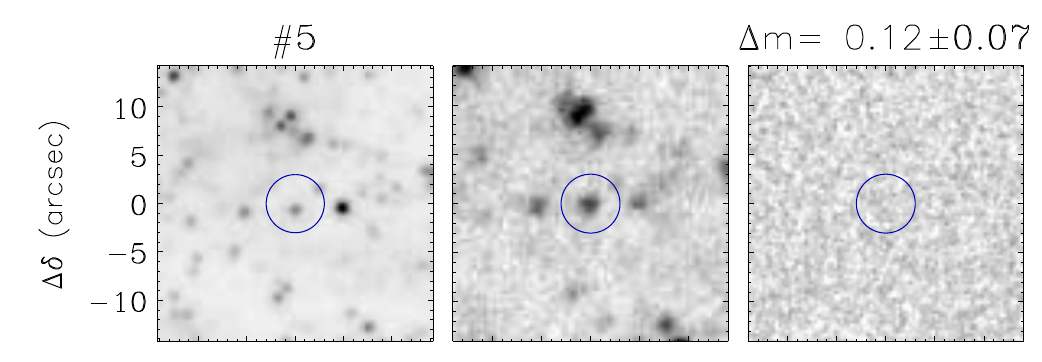}
}
\centerline{
\includegraphics[width=0.5\textwidth]{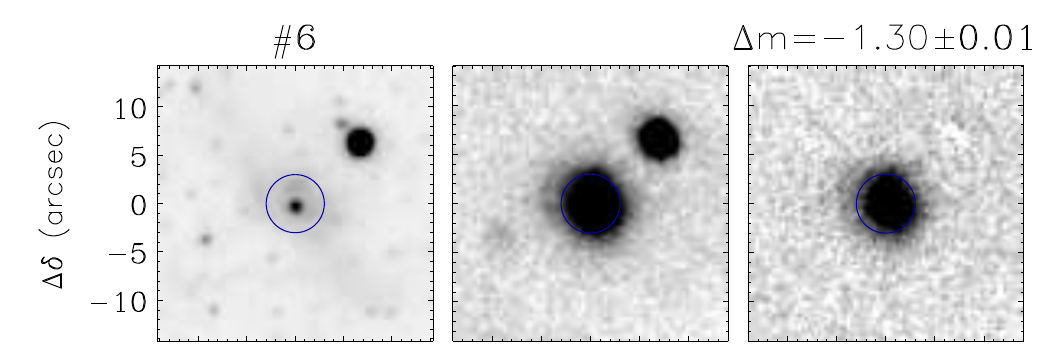}
\includegraphics[width=0.5\textwidth]{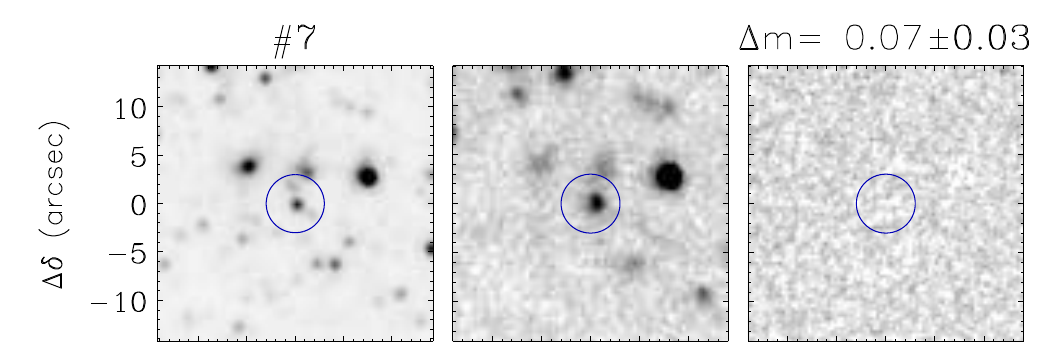}
}
\centerline{
\includegraphics[width=0.5\textwidth]{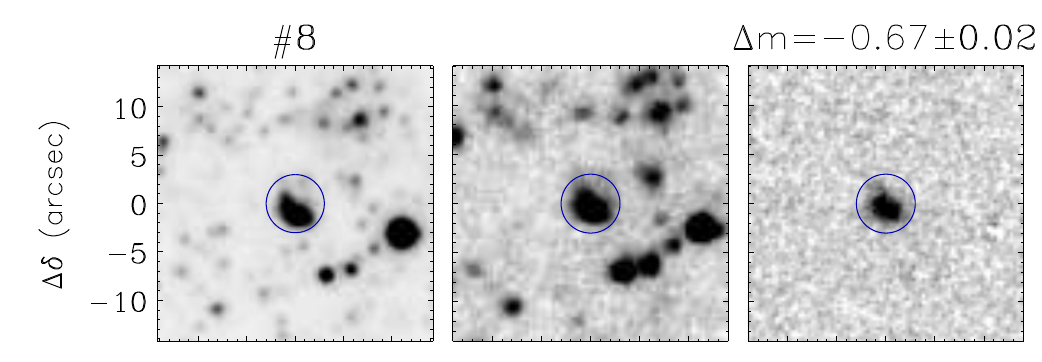}
\includegraphics[width=0.5\textwidth]{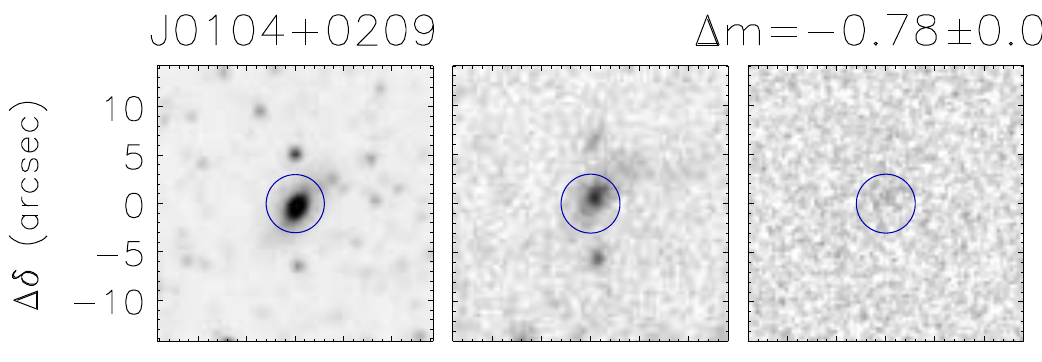}
}
\caption{
$15''\times15''$ regions around candidates to  WR-stars listed according their numbers in \citet{ArmandroffMassey1985} and the background galaxy J010457.6+020953. From left to right: $r$-band DESI Legacy survey image \citep{Legacy2019AJ....157..168D}, MaNGaL images in the filter centered on the \HeII{} emission line before and after continuum subtraction. Blue circle   (6 arcsec in diameter) is centered on the considered object. The $\Delta m$ is  a difference between  magnitudes in the filters centered on the line and continuum.
}.
\label{fig:HeII_WR}
\end{figure*}

\subsection{Search for WR candidates in MUSE spectral data}

We identified three out of the eight candidates (\#3, \#4, \#5) mentioned in the work of \citet{ArmandroffMassey1985} within the MUSE fields, and extracted their spectra. \revone{According to spectral data, source \#4 is a group of stars of similar early spectral class types, while sources \#3 and \#5 are single stars of the early spectral class.} 

We first performed reddening correction to the spectra using the \textsc{pyneb} package \citet{Luridiana2015} and the parametrization of the reddening curve from the work of \citet{1999PASP..111...63F}. Since absorption lines in the spectra of stars are superimposed on the emission of the surrounding \HII{} complex, we used the spectra of regions close to the studied stars extracted from the MUSE data.

To estimate effective temperatures of the studied stars we then used the spectral absorption line of \Hb, which is the only apparent stellar detail in the spectra, comparing it with the stellar models. Given the strong emission details observed from the surrounding complex \HII{} in all spectra, we subtracted the \Hb{} emission component from each spectrum, performing \Hb{} and \Ha{} emission and absorption modelling simultaneously for each of the temperature models. We used SynthV \citep{2019ASPC..518..247T} program and ATLAS9 \citep{2003IAUS..210P.A20C} stellar atmosphere models (for the metallicity [Fe/H]=0.0 and the gravity log~$g$=4.5) to calculate the synthetic spectra of the stars. To estimate the effective temperatures of the stars, their spectra were normalized to the continuum level, and the synthetic spectra were convolved with an instrumental profile corresponding to the resolution of the MUSE spectrograph.

According to our modeling, all of the stars constituing the object \#4 from \citet{ArmandroffMassey1985} are well described by the model of the star with $T_{eff}\approx30kK$. The temperature  for the source \#5 can not be estimated  by this method due to the significantly larger contribution from the nebula radiation for the \Hb{} line. Based on the available spectrum, we can only conclude that the star belongs to the early spectral type showing no details characteristic for evolved objects. Plots related to our analysis are presented in \ref{sec:appendixB} on Fig.~~\ref{fig:MUSE}.

Third cWR star (object \#3 in \citealt{ArmandroffMassey1985}) shows a broad emission feature in the Ha line (FWHM=9.6\AA\; comparing to FWHM$\approx$3\AA \; for nebula lines for the MUSE data on the same wavelengths, Gaussian fitting is presented on Fig.~\ref{fig:MUSE}). The image does not show any further detail that would clearly indicate it's nature, but such a wide \Ha{} line indicates that the star is losing a large amount of its matter (for example, through the stellar wind). 

The spectrum of another WR star candidate, cWR \#2, was obtained with a long slit. The object hits the slit PA=15 (see Fig.~\ref{fig:Ha_map}), the full spectrum is shown in Fig.~\ref{fig:sp_cWR2}. The spectrum demonstrates absorbtion lines in Balmer series and \HeI$\lambda$4471 line, which is characteristic of an early type star, but also exhibits an emission component in the \Ha{} line. The FWHM of the line does not exceed the spectral resolution of the image in this region. Temperature estimation base on absorption lines gives $\mathrm{24kK}$. We attribute the star to the B1e type. 

\section{\Ha{} emission stars from MUSE data}
\label{emission_stars_fromMUSE}

MUSE archival data turned out to contain a large number of point radiation sources in the \Ha{} line. We mapped 32 \Ha-emission stars by extracting the spectra of all \Ha{} sources from the available MUSE fields and visually analyzing them (see Fig.~\ref{MUSE_Ha_point_sources}). Coordinates of the sources are listed in Appendix \ref{sec:appendixC}.

To sample the stellar spectra, we used circular apertures that included the entire \Ha{} emission region. We subtracted the background from the obtained spectra, and fitted the \Ha{} line with a Gaussian function. We selected sources with S/$N_\mathrm{H{\alpha}}$ > 15, and also checked that selected spectra show a continuum, but do not show lines characteristic of \HII{} regions (\SII{}, \NII{}, \OIII{} lines).

Most of the spectra did not show any unusual spectral features except \Ha{} emission. However, we found two sources showing the P Cyg profile. We present their spectra on Fig.~\ref{MUSE_PCyg}. 

One of the objects was previously described in the work of \citet{2007ApJ...671.2028B}, including the only optical spectrum received. The object type was defined as A2 Ia. Photometry for the star was also published in the works of \citet{2015MNRAS.452..910M}, \citet{2020yCat.1350....0G} and \citet{Garcia2009A&A...502.1015G}. Using V magnitude from SIMBAD database (\citealt{2000A&AS..143....9W}) from the work of \citet{Garcia2009A&A...502.1015G}, $R_V$=3.1 and $\mathrm{E}_{\mathrm{B-V}}$=0.036, we estimate $M_\mathrm{V}$=-8.04.
 
The spectrum of this star has lines of \CaII$\lambda$8498,8542,8662, which indicate the presence of a rarefied nebula around the object and are the characteristic spectral details of warm hypergiants (see \citealt{2017ApJ...836...64H}). Absorption lines of the Paschen series are represented, as well as \FeII$\lambda$4924,5018,5169,5235,5276,5317, \OI$\lambda$7773,8446 and \SiII$\lambda$6347,6371 lines. The spectrum is similar to the spectra of warm hypergiants J004507.65+413740.8, J004621.08+421308.2 of the M33 galaxy studied in \citet{2017ApJ...836...64H}, \citet{2020MNRAS.497..687S}. Comparing  with previously studied spectra of LBV stars in the M33 galaxy, a strong similarity is revealed with the spectra of V532 \citet{1978A&A....67..291R} and Var C at the time of their maximum brightness (see \citealt{2011AstBu..66..123S, 2013ATel.5538....1V, 2015A&A...581A..12B}). Hypergiants normally do not show spectral variability, therefore it is necessary to examine the star for variability to finally classify it.

Another star is not described in the literature. No \CaII\, or [Ca II] lines are observed. In general, the spectrum is similar to the spectrum of the star J004051.59+403303.0 from the M31 galaxy (\citealt{2015MNRAS.447.2459S}). By analogy, it can be classified as a candidate to LBV star at its maximum brightness or a B[e] supergiant.

\begin{figure}
\includegraphics[scale=0.70]{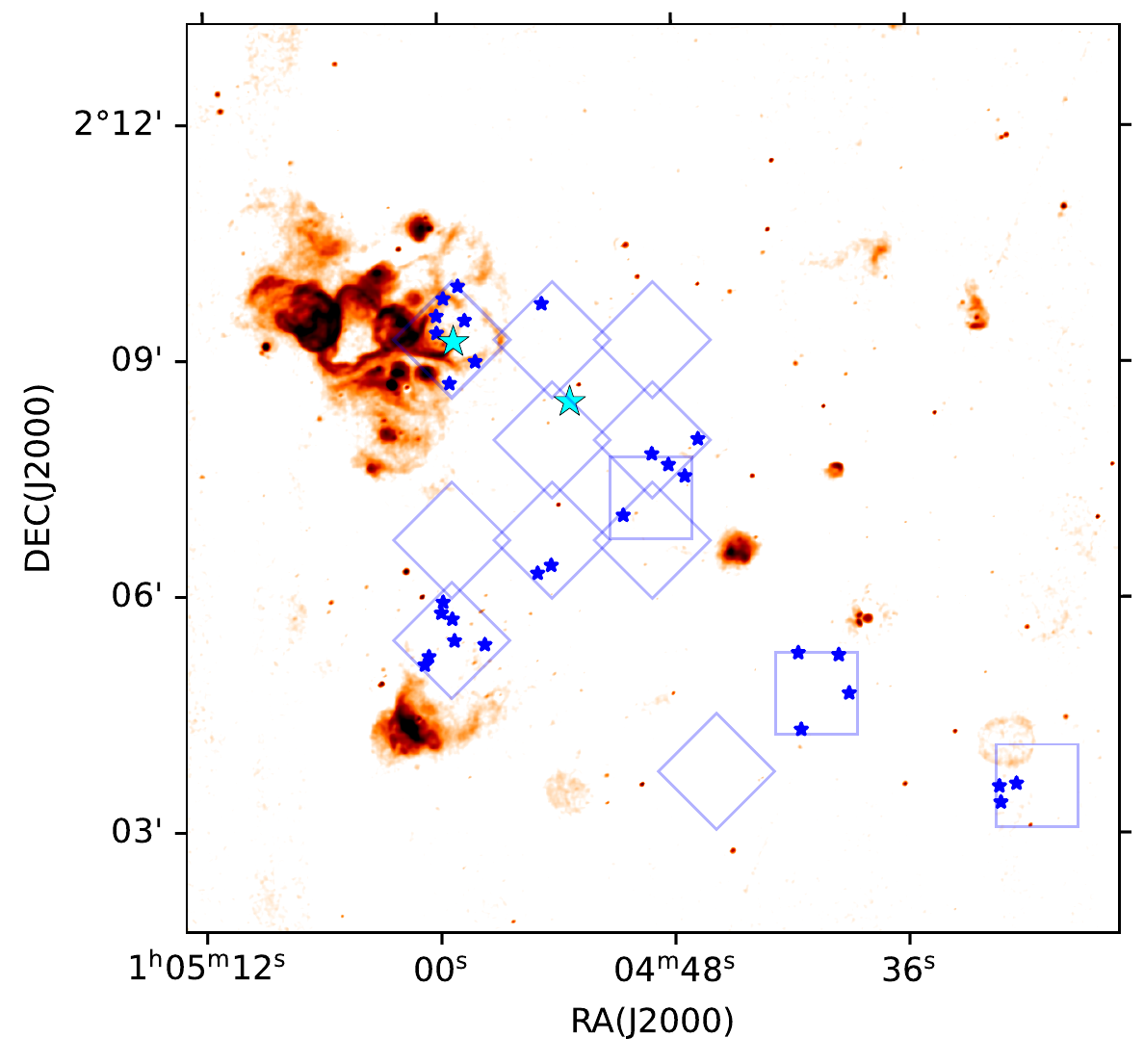}
\caption{Point source analysis: blue stars point \Ha{} emission point sources from MUSE archive data; Positions of two PCyg-profile stars found between \Ha{} emission stars is showed by cyan stars. The underlying map is the \Ha\ image from 1.5-m telescope.
}
\label{MUSE_Ha_point_sources}
\end{figure}

\begin{figure*}
\includegraphics[scale=0.93]{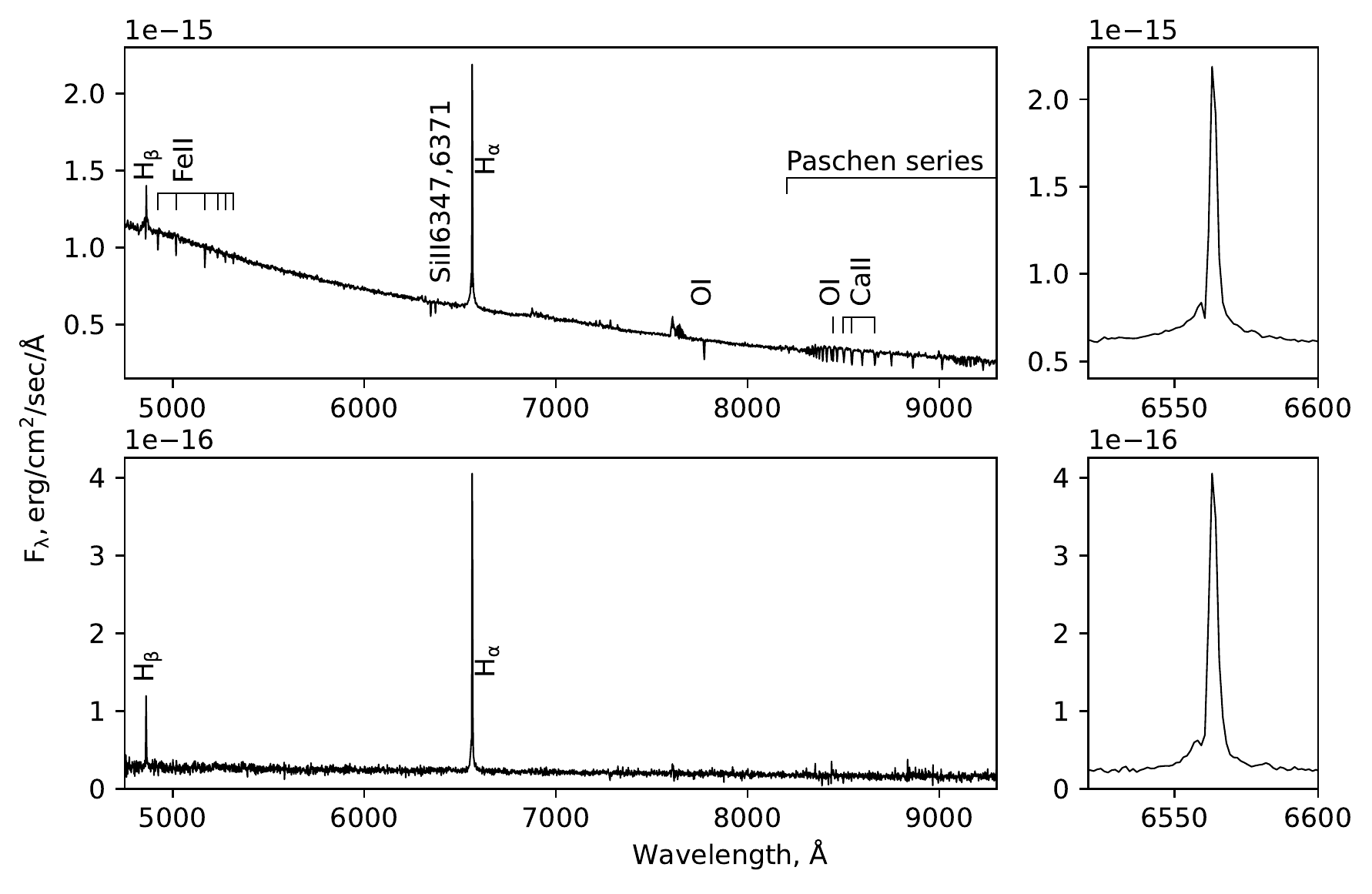}
\caption{PCyg-profile stars found in MUSE archive data.}

\label{MUSE_PCyg}
\end{figure*}

\section{Discussion}

\subsection{Faint \HII\ regions}

We conducted a study of faint ionised nebulae in the dwarf galaxy IC 1613 based on narrow-band imaging in \Ha, \HeII{} and \SII{} lines. The obtained images are the deepest to date, which made it possible to reveal several new faint structures in the \HII{} regions of the galaxy. \revone{Several of these shells (B1--B3 and B7) are located in the regions of dense neutral hydrogen, while other shells (B4--B6) lie near the edges of the \HI{} areas}. For the detected \revone{faint shells} of ionised gas, we carried out new spectral observations. We analysed these nebular regions using the BPT diagnostic diagrams and calculated the energy balance between the radiation of the found \revone{shells} and the stars located in the centre of the detected \revone{shells}. 

The BPT diagrams revealed that none of the \revone{shells} show clear signatures of shock excitation based on the spectra of the \Ha{} regions $S_3$--$S_{10}$. \revone{The $S_2$ region, associated to the B1 shell, is located on the BPT diagram separately from the rest, lying only slightly below the maximum starburst line and also far from the photoionization model grid for the metallicity of IC1613, which indicates the possible contribution of shocks to the ionisation of the gas.} \revone{Shells B1, B3 and B7} are most likely ionised  by the OB associations in their centers, according to the positions of massive stars from the catalogue of \citet{Garcia2009A&A...502.1015G}. 

In the case of the B2 \revone{shell}, the BPT diagrams, together with the calculated energy balance, raise a question about the possible nature of this object. Optical data show no stars of early spectral classes that could release the required amount of energy into the ISM inside the \revone{shell}. Radio observations by \citet{2022AJ....163...66S} show no detection of any sources within it. However, based on the data from XMM-{\it Newton} and {\it Swift}/XRT, we discovered an X-ray source with a 10" offset from the center of the \revone{shell}. Assuming that the source is located in the galaxy IC 1613, its intrinsic luminosity is $L_{X}\approx 6\times10^{36}$ erg s$^{-1}$, which is characteristic value for X-ray binaries and pulsars surrounded with nebulae. The B2 \revone{shell}, lacking central clusters as well as single O stars but exhibiting comparable \Ha{} flux to B1 and B3 regions together with X-ray emission, possibly connected to the \revone{shell}, presents a potential SNR. However, its classification remains inconclusive. Examples of the SNR in metal poor environment demonstrating no bright counterparts except X-ray emission could be found in the literature. For example, supernova exploded in low-density Galactic halo environment was recently found in the SRG/eROSITA all-sky survey and is described in the work of \citet{Churasov_SNR}. 


Among the spectra of faint regions in the galaxy IC 1613, one stands out as a bright \OIII{} line source (spectrum $S_8$, for which \OIIIHb\ $> 8.8$). Such line ratio can be explained by assuming the presence of a planetary nebula.

\subsection{Emission stars in IC 1613}

\revone{In the work of \citet{ArmandroffMassey1985}, 8 WR candidates were listed for IC 1613 galaxy, 5 of them extremely probable.} Our search for the WR stars in IC 1613,  \revone{based on emission in \HeII{} line, does not confirm any of these, except for one previously known WO star (object \#6). Moreover, it does not reveal any additional new WR star candidates.} 
\revone{The findings for the four of the 8 WR candidates from \citet{ArmandroffMassey1985} are supported from the analysis of the archival MUSE data for the objects \#3,\#4,\#5 and long-slit BTA spectroscopy for object \#2. None of them proved to be a WR star. We discuss further each of these four stars.}

\revone{Based on our estimates, the temperatures of the group of stars corresponding to the position of the object \#4 is close to 30 kK, whereas for the star \#5, the temperature could not be measured due to significant emission from the surrounding \HII{} complex. The star probably belongs to an early spectral class, since the spectrum does not show any signs of line emission characteristic for the evolved stars. }
Star \#3 demonstrates a broad \Ha{} line (FWHM = 9.6 \AA), indicating the loss of the stellar matter. No other notable spectral features are seen. Some examples of similar objects are presented in the literature. In the work of \citet{2007AJ....134.2474M}, which aimed at searching for new LBV star candidates, examples of three \Ha{} emission-line stars are given, which do not show any other spectral features. One of the objects also exhibits photometric variability at the $0.7$ mag level in a 10-year long period (J002016.48+591906.9 in dwarf galaxy IC 10). \revone{Thus, for the star \#3 variability should be searched.}
\revone{ The long-slit BTA spectrum for the star \#2 shows \Ha{} emission together with absorption in the other Balmer series lines and the \HeI$\lambda$4471 line. From our estimated temperature of $T=24kK$ we classify it as a B1e star.}

\revone{However}, among these four WR candidates with available spectra we observed strong \Ha{} emission associated with the two of them \revone{(\#2 and \#3), whereas the \Ha{} emission seen in the spectra of candidates \#4 and \#5 is related to the surrounding \HII{} region (Fig.~\ref{fig:MUSE}).} 

\revone{Thus, we suggest that the galaxy IC 1613 does not exhibit a high rate of WR star formation, as previously discussed.}
In any case this high fraction of the emission stars (2 out of 4)  cannot be accidental. Preliminary we can assume that the stars \#2 and \#3 have undergone a transient helium emission more than 35 years ago when the observations by \citet{ArmandroffMassey1985} were performed. Whereas the wrong detection  of the helium line for candidates \#4 and \#5 could be caused by underestimated continuum contribution in the narrow-band  filter corresponding to \HeII{} in these blue stars. 

\revone{Finally, our investigation of archival} MUSE data revealed 32 additional stars with \Ha{} emission. Two stars from the sample demonstrate P Cyg profiles in \Ha{} and \Hb{} lines. One of the stars, previously described in the work of \citet{2007ApJ...671.2028B}, has $M_V=-8.04$ and contains Balmer and Paschen series, \CaII, \SiII, \OI, \FeII{} lines in the spectrum. These spectral properties point to a warm hypergiants, but could also indicate a spectrum of an LBV star V532 (Romano star) and Var C. To accurately classify this star, it should be further followed and checked for spectral variability. The other star is less luminous, demonstrating only \Ha{} and \Hb{} emission. We classify it as a candidate to LBV star at its maximum brightness or a B[e] supergiant.

\section{Conclusion}


Using presently the deepest mapping in the \Ha{} and \HeII{} emission lines of the ionised gas in the nearby dIrr galaxy IC~1613 together with new long-slit spectroscopy at the 6-m BTA and archival 8-m MUSE/VLT data we carried out a study of faint ionised \revone{regions} and emission from massive stars in this galaxy. The main findings are the following:

\begin{itemize}
    \item The most of the \revone{shells} (B1, B3--B7) are related with OB associations. Their ionisation is caused by photoionisation of UV photons of nearest O-stars, as it followed from their line ratio diagnostic diagrams and energy budget evaluation (\revone{the case of B1, B3 and B7}).
    \item The B2 \revone{shell} is intriguing as there are no stars that could release the required UV-luminosity to ionise it. \revone{There is the} X-ray source projected into the \revone{shell}, however it may be also interpreted as a background galaxy. Preliminary we consider B2 as a candidate of a peculiar SNR, however new multi-wavelength observations are required. 
  \item We have found a new PN-like nebula between B6 and B7 \revone{shells}, which corresponds to \revone{the} spectrum $S_8$.
  \item \revone{We rejected 5 studied WR candidates with \HeII{} photometry and additionally confirm the results with spectroscopy for 4 of them (\#2, \#3, \#4, \#5). WR candidates \#2 and \#3 show strong \Ha{} emission.} 
\item We analysed spectra of 32 stars with  \Ha{} emission from MUSE data, two of them also demonstrate PCyg profile, both of which could potentially belong to LBV stellar type.  
   
\end{itemize}

\revone{In summary, we suggest that the faint H$\alpha$ \revone{shells} in nearby dwarf galaxy IC 1613 are powered by OB association or individual stars, except for one pointing to possible shock wave ionization and SNR origin. We discuss that the galaxy has lower rate of WR stars, as previously reported.  }


\section*{Acknowledgements}
The authors thank M. Garcia for providing data from the photometric catalog of IC 1613 stars. We also thank T. Lozinskaya, L. Oskinova, A. Smirnova, K. Vasiliev \revone{and M. Filipovi{\'c}} for useful discussion and advices. We express our gratitude to all telescope observers (D. Oparin, E. Malygin, A. Grokhovskaya, S. Dodonov, U. D. G{\"o}ker). \revone{We thank the referee for useful comments and suggestions.} 

We obtained part of the observed data on the unique scientific facility `Big Telescope Alt-azimuthal' of SAO RAS as well as analysed the spectral and imaging data  with the financial support of grant No~075-15-2022-262 (13.MNPMU.21.0003) of the Ministry of Science and Higher Education of the Russian Federation. Observations with 1.5-m RTT were obtained through Postdoc Scholarship of  Bosphorus University (project No~14B03P4).

M.M.V. and D.I. acknowledge funding provided by the University of Belgrade - Faculty of Mathematics (the contract No~451-03-47/2023-01/200104) through the grant by the Ministry of Science, Technological Development and Innovation of the Republic of Serbia.

The Legacy Surveys consist of three individual and complementary projects: the Dark Energy Camera Legacy Survey (DECaLS; Proposal ID \#2014B-0404; PIs: David Schlegel and Arjun Dey), the Beijing-Arizona Sky Survey (BASS; NOAO Prop. ID \#2015A-0801; PIs: Zhou Xu and Xiaohui Fan), and the Mayall z-band Legacy Survey (MzLS; Prop. ID \#2016A-0453; PI: Arjun Dey). DECaLS, BASS, and MzLS together include data obtained, respectively, at the Blanco telescope, Cerro Tololo Inter-American Observatory, NSF’s NOIRLab; the Bok telescope, Steward Observatory, University of Arizona; and the Mayall telescope, Kitt Peak National Observatory, NOIRLab. NOIRLab is operated by the Association of Universities for Research in Astronomy (AURA) under a cooperative agreement with the National Science Foundation.

This project used data obtained with the Dark Energy Camera (DECam), constructed by the Dark Energy Survey (DES) collaboration.

This project used data obtained with the Dark Energy Camera (DECam), which was constructed by the Dark Energy Survey (DES) collaboration. Funding for the DES Projects has been provided by the U.S. Department of Energy, the U.S. National Science Foundation, the Ministry of Science and Education of Spain, the Science and Technology Facilities Council of the United Kingdom, the Higher Education Funding Council for England, the National Center for Supercomputing Applications at the University of Illinois at Urbana-Champaign, the Kavli Institute of Cosmological Physics at the University of Chicago, Center for Cosmology and Astro-Particle Physics at the Ohio State University, the Mitchell Institute for Fundamental Physics and Astronomy at Texas A\&M University, Financiadora de Estudos e Projetos, Fundacao Carlos Chagas Filho de Amparo, Financiadora de Estudos e Projetos, Fundacao Carlos Chagas Filho de Amparo a Pesquisa do Estado do Rio de Janeiro, Conselho Nacional de Desenvolvimento Cientifico e Tecnologico and the Ministerio da Ciencia, Tecnologia e Inovacao, the Deutsche Forschungsgemeinschaft and the Collaborating Institutions in the Dark Energy Survey. The Collaborating Institutions are Argonne National Laboratory, the University of California at Santa Cruz, the University of Cambridge, Centro de Investigaciones Energeticas, Medioambientales y Tecnologicas-Madrid, the University of Chicago, University College London, the DES-Brazil Consortium, the University of Edinburgh, the Eidgenossische Technische Hochschule (ETH) Zurich, Fermi National Accelerator Laboratory, the University of Illinois at Urbana-Champaign, the Institut de Ciencies de l’Espai (IEEC/CSIC), the Institut de Fisica de Altes Energies, Lawrence Berkeley National Laboratory, the Ludwig Maximilians Universitat Munchen and the associated Excellence Cluster Universe, the University of Michigan, NSF’s NOIRLab, the University of Nottingham, the Ohio State University, the University of Pennsylvania, the University of Portsmouth, SLAC National Accelerator Laboratory, Stanford University, the University of Sussex, and Texas A\&M University.

BASS is a key project of the Telescope Access Program (TAP), which has been funded by the National Astronomical Observatories of China, the Chinese Academy of Sciences (the Strategic Priority Research Program ‘The Emergence of Cosmological Structures’ Grant \#XDB09000000), and the Special Fund for Astronomy from the Ministry of Finance. The BASS is also supported by the External Cooperation Program of Chinese Academy of Sciences (Grant \# 114A11KYSB20160057), and Chinese National Natural Science Foundation (Grant \# 11433005).

The Legacy Survey team makes use of data products from the Near-Earth Object Wide-field Infrared Survey Explorer (NEOWISE), which is a project of the Jet Propulsion Laboratory/California Institute of Technology. NEOWISE is funded by the National Aeronautics and Space Administration.

The Legacy Surveys imaging of the DESI footprint is supported by the Director, Office of Science, Office of High Energy Physics of the U.S. Department of Energy under Contract No. DE-AC02-05CH1123; by the National Energy Research Scientific Computing Center, a DOE Office of Science User Facility under the same contract; and by the U.S. National Science Foundation, Division of Astronomical Sciences under Contract No. AST-0950945 to NOAO.
 
\section*{Data Availability}

The data underlying this article will be shared on reasonable request to the corresponding author.



\bibliographystyle{mnras}
\bibliography{Article} 



\appendix
\section{Spectra of the nebulae from long-slit spectroscopy data}\label{sec:appendixA}

In this section, we present long-slit emission spectra of faint nebulae obtained with a 6-meter BTA telescope (Fig.~\ref{Spectra_S1-S5} and \ref{Spectra_S6-S10}). Results based on the listed spectra are described in sections \ref{sec:diff_ion_shells}.

\begin{figure*}
\includegraphics[scale=0.8]{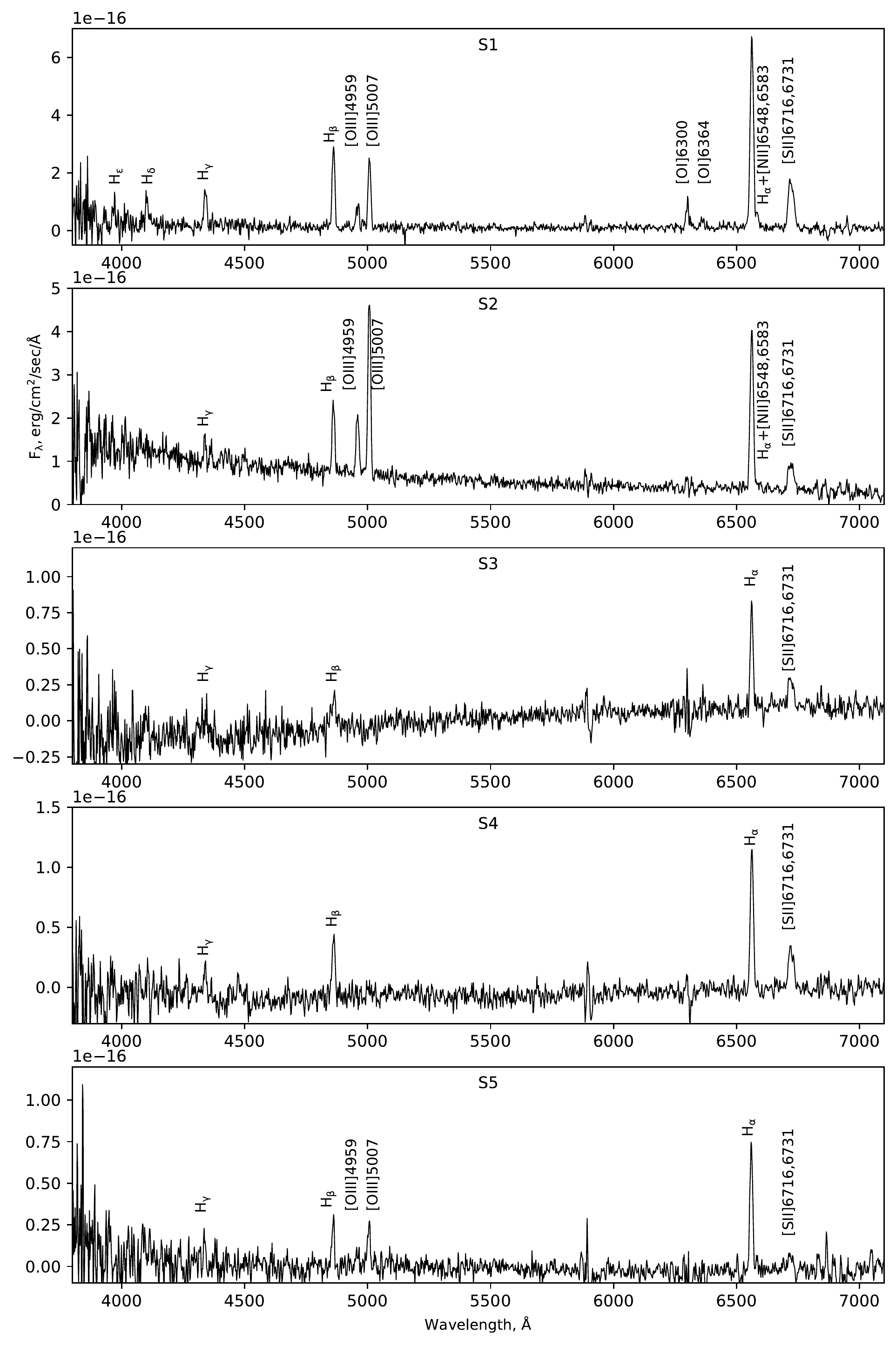}
\caption{Emission spectra S1-S5 \revone{of faint \HII{} regions B1-B3} obtained on 6-m BTA telescope. Fluxes corresponding to emission lines are listed in the table \ref{tab:fluxes}. Spectra are numerated as on Fig.~\ref{fig:Ha_map}}
\label{Spectra_S1-S5}
\end{figure*}

\begin{figure*}
\includegraphics[scale=0.8]{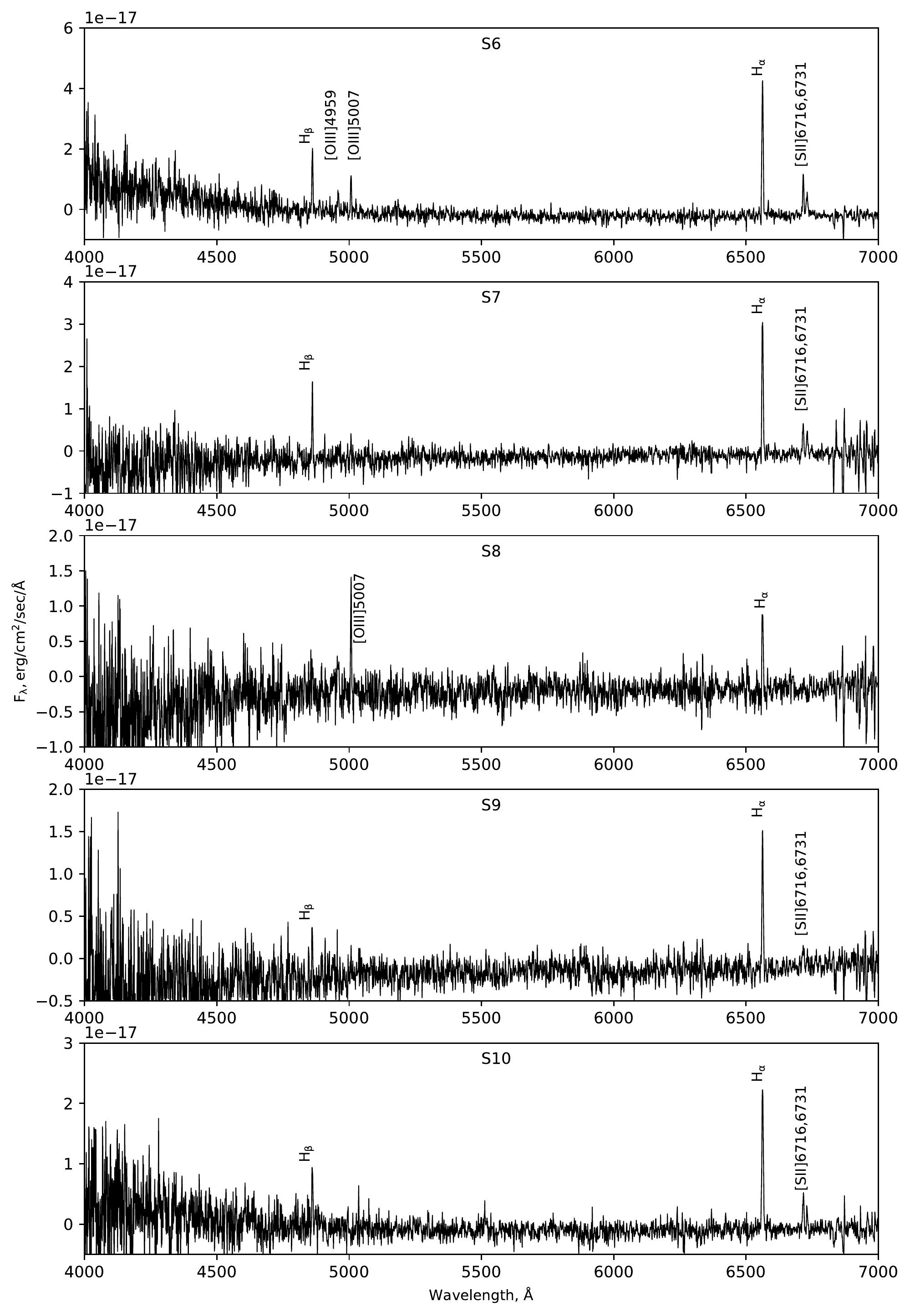}
\caption{Emission spectra S6-S10 \revone{of faint \HII{} regions B4-B7} obtained on 6-m BTA telescope.}
\label{Spectra_S6-S10}
\end{figure*}

\section{Spectra of candidates to WR stars}\label{sec:appendixB}

In this section, we present the spectra of candidates to WR stars, listed in the work of \citet{ArmandroffMassey1985}. We found the spectra of the three listed candidates in the archived MUSE data (stars cWR $\#3,\#4,\#5$, see Fig.~\ref{fig:MUSE}), and one spectrum we obtained from our new observation with the 6-m BTA Telescope (cWR $\#2$, see Fig.~\ref{fig:sp_cWR2}). None of the studied stars showed signs of a WR star. The results are described in detail in the section \ref{sec:WR_candidates}.

\begin{figure*}
\includegraphics[scale=0.9]{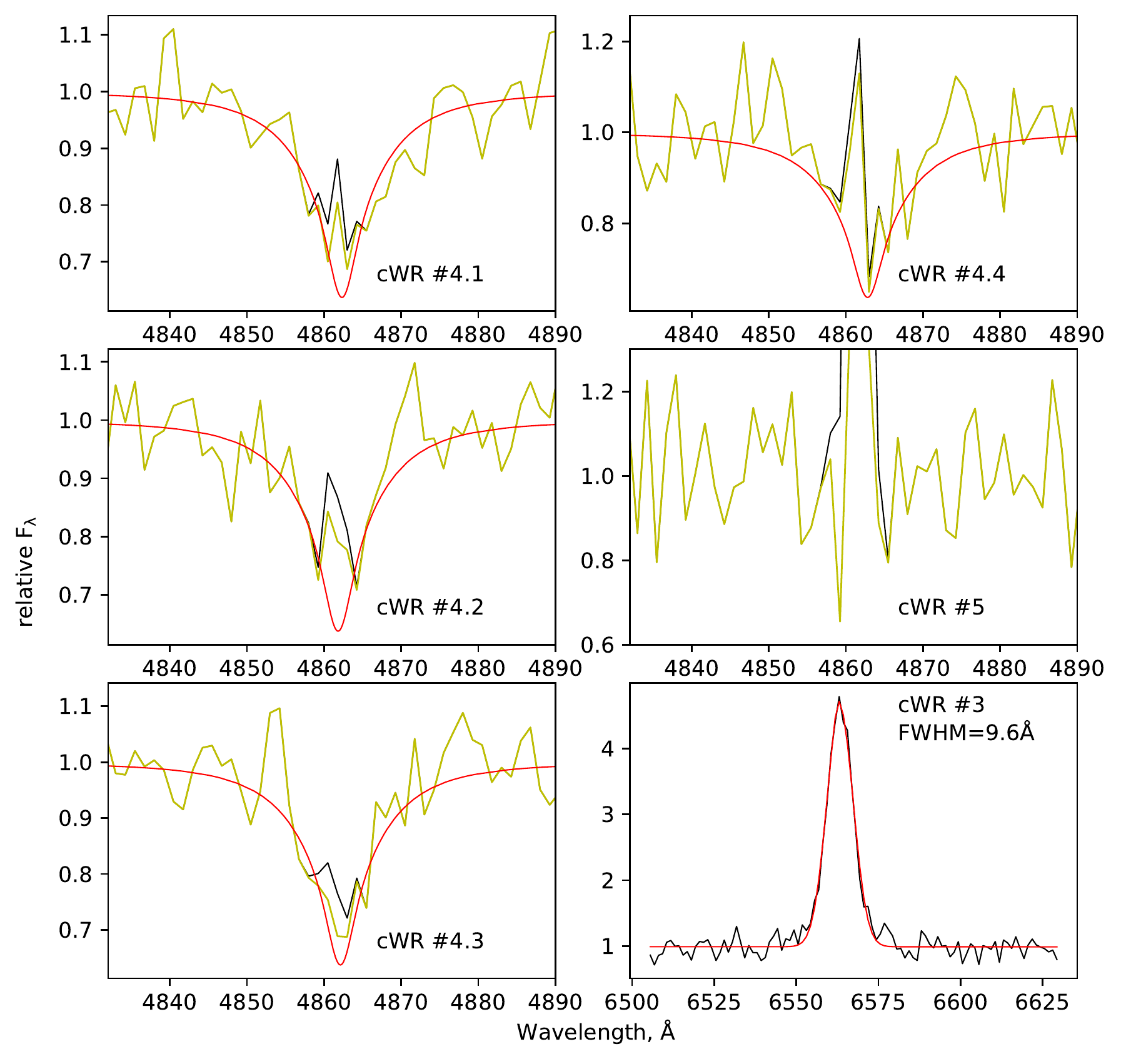}
\caption{\Hb{} and \Ha{} lines extracted from spectra obtained from MUSE  data cube 
for candidates to WR stars $\#3, \#4, \#5$, as listed in the work of \citet{ArmandroffMassey1985}. First five spectra: black line shows the real \Hb{} profile normalized to the continuum level. Yellow line shows the line profile after subtracting the emission from the surrounding \HII{} complex. The red line shows the line profile model for a star with an effective temperature of 30 kK. Object \#4 is resolved into a group of four early-type stars with effective temperatures $\sim$30kK. Object $\#5$ is an early-type star, but it is not possible to reliably estimate it's temperature due to the strong emission of the \HII{} complex in the area where the star is located. Lower right spectrum: black line shows \Ha{} profile from the star. Modeling (red line) shows that there is no emission nebular components but only a wide emission from the star. For a description of the simulation and a discussion, see the section \ref{sec:WR_candidates}.
}
\label{fig:MUSE}
\end{figure*}

 \begin{figure*}
\includegraphics[scale=0.93]{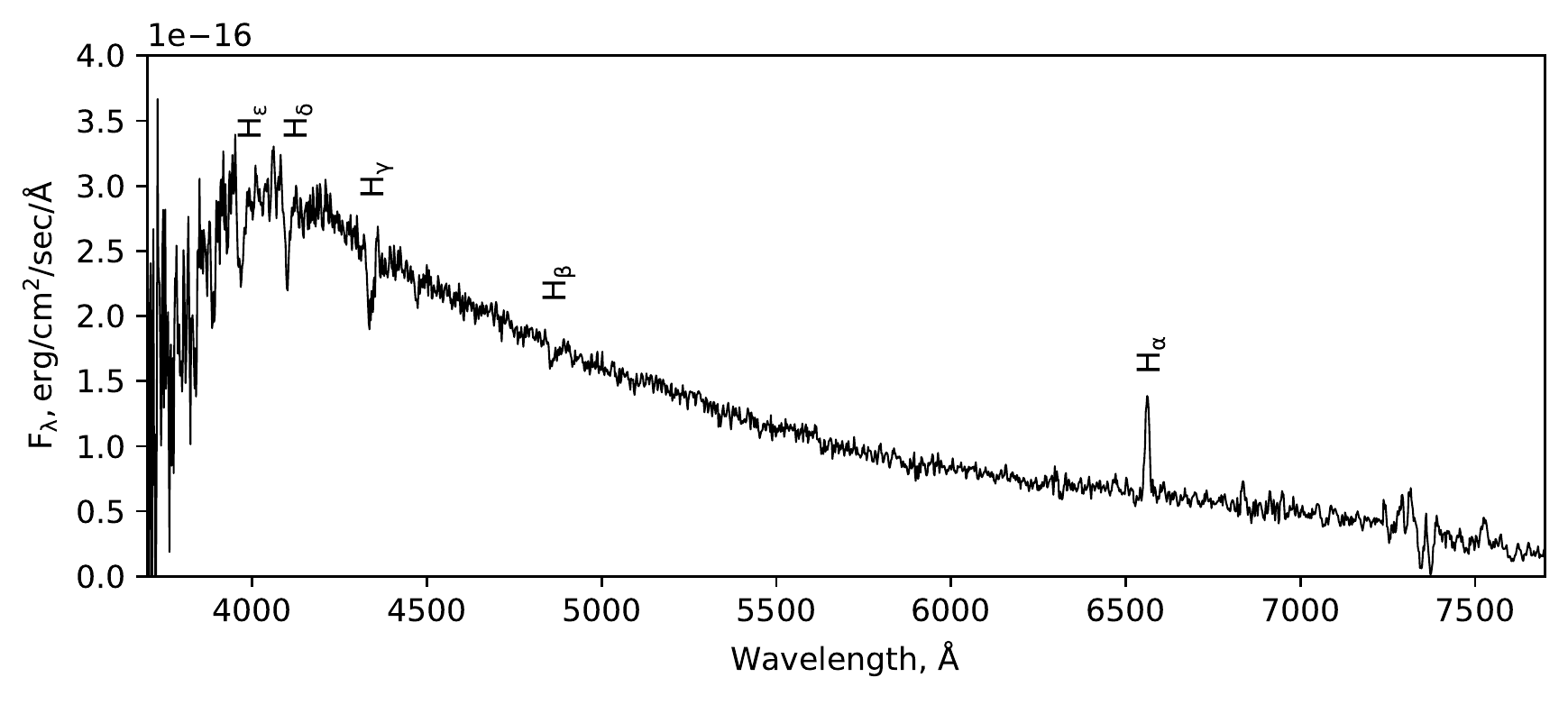}
\caption{Long-slit spectrum (PA=15) of cWR $\#2$ star.}
\label{fig:sp_cWR2}
\end{figure*}

\section{Coordinates of stars showing \Ha \; emission}
\label{sec:appendixC}

We list in Table \ref{tab:coordinates} coordinates of the 32 stars from the MUSE data showing emission in \Ha{} line. 

\begin{table*}
\centering
\caption{Coordinates of 32 stars from MUSE data showing \Ha\; emission.}
\hspace{1.0cm}
\label{tab:coordinates}
\begin{tabular}
{|c|c|c|c|}
\hline
\multicolumn{4}{c}{RA DEC, hms dms}\\
\hline
01h04m58.1s +02d08m59.6s&
01h04m59.3s +02d09m14.8s&
01h04m59.4s +02d09m14.4s&
01h04m59.5s +02d08m42.9s\\
01h05m00.1s +02d09m34.2s&
01h05m00.1s +02d09m21.2s&
01h04m58.7s +02d09m31.0s&
01h04m59.0s +02d09m57.1s\\
01h04m59.8s +02d09m47.5s&
01h04m54.7s +02d09m43.8s&
01h04m53.3s +02d08m29.0s&
01h04m49.1s +02d07m49.1s\\
01h04m48.3s +02d07m40.9s&
01h04m46.7s +02d08m00.6s&
01h04m55.0s +02d06m18.1s&
01h04m54.3s +02d06m24.2s\\
01h04m50.6s +02d07m02.3s&
01h04m57.7s +02d05m23.7s&
01h05m00.8s +02d05m08.1s&
01h05m00.6s +02d05m14.5s\\
01h04m59.3s +02d05m26.6s&
01h04m59.4s +02d05m43.1s&
01h04m59.9s +02d05m47.7s&
01h04m59.8s +02d05m56.0s\\
01h04m39.6s +02d05m15.7s&
01h04m41.7s +02d05m17.2s&
01h04m41.5s +02d04m18.7s&
01h04m39.0s +02d04m46.5s\\
01h04m30.5s +02d03m37.4s&
01h04m31.4s +02d03m35.3s&
01h04m31.3s +02d03m23.0s&
01h04m47.4s +02d07m32.2s\\

\hline  
\end{tabular}
\end{table*}

 \bsp	
\label{lastpage}
\end{document}